\documentclass{sig-alternate}

\usepackage[margin=1in]{geometry}

\usepackage[utf8]{inputenc} 
\usepackage{algorithm}
\usepackage{algpseudocode}
\usepackage{graphicx}
\usepackage{textcomp}
\usepackage{xcolor}
\usepackage{listings}

\usepackage{scalerel}

\usepackage{subfigure}

\usepackage{tikz}
\usetikzlibrary{arrows}
\usetikzlibrary{backgrounds}
\usetikzlibrary{matrix}
\usetikzlibrary{positioning}
\usetikzlibrary{svg.path}
\usetikzlibrary{shapes}

\usepackage{adjustbox}
\usepackage{bbding}
\usepackage{enumitem}
\usepackage[T1]{fontenc}
\usepackage{upquote}
\usepackage{amsmath}

\usepackage{tabularx}
\usepackage{booktabs}
\usepackage{makecell}

\usepackage[htt]{hyphenat}

\input{definitions}

\usepackage{hyperref}

\makeatletter

\renewcommand*\env@matrix[1][*\c@MaxMatrixCols c]{%
  \hskip -\arraycolsep
  \let\@ifnextchar\new@ifnextchar
  \array{#1}}

\makeatother

\begin{document}

\pagestyle{empty}


\title{An analysis of the graph processing landscape}

\vspace{-20pt}
\author{Miguel E. Coimbra$^{*}$, Alexandre P. Francisco and Luís Veiga\\
\texttt{miguel.e.coimbra@tecnico.ulisboa.pt, aplf@inesc-id.pt, luis.veiga@inesc-id.pt}\\
INESC-ID/IST, Universidade de Lisboa, Portugal\\
\textit{*Corresponding author}\\
\vspace{-20pt}
}

\vspace{-20pt}

\maketitle

\begin{abstract}

The value of graph-based big data can be unlocked by exploring the topology and metrics of the networks they represent, and the computational approaches to this exploration take on many forms. 
For the use-case of performing global computations over a graph, it is first ingested into a graph processing system from one of many digital representations. 
Extracting information from gra\-phs involves processing all their elements globally, and can be done with single-machine systems (with varying approaches to hardware usage), distributed systems (either homogeneous or heterogeneous groups of machines) and systems dedicated to high-\-per\-formance computing (HPC).
For these systems focused on processing the bulk of graph elements, common use-cases consist in executing for example algorithms such as PageRank or community detection, which produce insights on graph structure and relevance of their elements.

Considering another type of use-case, graph-specific databases may be used to efficiently store and represent graphs to answer requests like queries about specific relationships and graph traversals.
While tabular-type databases may be used to store relations between elements, it is highly inefficient to use for this purpose these databases in terms of both storage space requirements and processing time.
Relational database management systems (RDBMS) and NoSQL databases, to achieve this purpose, need complex nested queries to represent the multi-level relations between data elements.
Graph-specific databases employ efficient graph representations or may make use of underlying storage systems.

In this survey we firstly familiarize the reader with common graph datasets and applications in the world of today.
We provide an overview of different aspects of the graph processing landscape and describe classes of systems based on a set of dimensions we describe.
The dimensions we detail encompass paradigms to express graph processing, different types of systems to use, coordination and communication models in distributed graph processing, partitioning techniques and different definitions related to the potential for a graph to be updated.
This survey is aimed at both the experienced software engineer or researcher as well as the newcomer looking for an understanding of the landscape of solutions (and their limitations) for graph processing.

\end{abstract}

\section{Introduction}\label{sec:introduction}

Graph-based data is found almost everywhere, with examples such as analyzing the structure of the Wor\-ld Wide Web~\cite{DBLP:conf/www/BoldiV04,DBLP:conf/dcc/BoldiV04,brin1998anatomy}, bio-informatics data representation via \textit{de Bruijn} graphs~\cite{ACombinatorialProblem1946} in metagenomics~\cite{Li01022010,18349386}, atoms and covalent relationships in chemistry~\cite{balaban1985applications}, the structure of distributed computation itself~\cite{Malewicz:2010:PSL:1807167.1807184}, massive parallel learning of tree ensembles~\cite{panda2009planet} and parallel topic models~\cite{smola2010architecture}.
Academic research centers in collaboration with industry players like Facebook, Microsoft and Google have rolled out their own graph processing systems, contributing to the development of several open-source frameworks~\cite{graphengine, ching2013scaling, graphx, carbone2015apache}.
They need to deal with huge graphs, such as the case of the Facebook graph with billions of vertices and hundreds of billions of edges~\cite{facebook_statistics2020}.

\subsection{Domains}\label{sec:introduction:sec:domains}

We list some of the domains of human activity that are best described by relations between elements - graphs:

\begin{itemize}

\item \textbf{Social networks.}
They make up a large portion of social interactions in the Internet.
We name some of the best-known ones: Facebook (2.50 billion monthly active users as of December 2019~\cite{facebook_newsroom}), Twitter (330 million monthly active users in Q1'19~\cite{twitter_investor_info}) and LinkedIn (330 million monthly active users as of December 2019~\cite{linkedin_statistics2020}).
In these networks, the vertices represent users and edges are used to represent friendship or follower relationships.
Furthermore, they allow the users to send messages to each other.
This messaging functionality can be represented with graphs with associated time properties. 
Other examples of social networks are WhatsApp (1.00 billion monthly active users as of early 2016~\cite{whatsapp_statistics2020}) and Telegram (300 million monthly active users~\cite{telegram_litigation2019}).


\item \textbf{World Wide Web.}
Estimates point to the existence of over 1.7 billion websites as of October 2019~\cite{total_number_websites}, with the first one becoming live in 1991, hosted at CERN.
Commercial, educational and recreational activities are just some of the many facets of daily life that gave shape to the Internet we know today.
With the advent of business models built over the reachability and reputation of websites (e.g. Google, Yahoo and Bing as search engines), the application of graph theory as a tool to study the web structure has matured during the last two decades with techniques to enable the analysis of these massive networks~\cite{DBLP:conf/www/BoldiV04,DBLP:conf/dcc/BoldiV04}.

\item \textbf{Telecommunications.}
These networks have been used for decades to enable distant communication between people and their structural properties have been studied using graph-based approaches~\cite{baritchi2000discovering,balasundaram2006graph}.
Though some of its activity may have transferred to the applications identified above as social networks, they are still relevant.
The vertices in these networks represent user phones, whose study is relevant for telecommunications companies wishing to assess closeness relationships between subscribers, calculate churn rates, enact more efficient marketing strategies~\cite{al2019social} and also to support foreign signals intelligence (SIGINT) activities~\cite{pfluke2019history}.

\item \textbf{Recommendation systems.}
Graph-based approaches to recommendation systems have been heavily explored in the last decades~\cite{grujic2008movies,gu2010collaborative,silva2010graph}.
Companies such as Amazon and eBay provide suggestions to users based on user profile similarity in order to increase conversion rates from targeted advertising.
The structures underlying this analysis are graph-based~\cite{zhao2017meta,yang2018graph,beyene2008ebay}.

\item \textbf{Transports, smart cities and IoT.}
Graphs have been used to represent the layout and flow of information in transport networks comprised of people circulating in roads, trains and other means of transport~\cite{euler1956seven,unsalan2012road,rathore2016exploiting}.
The Internet-of-Things (IoT) will continue to grow as more devices come into play and 5G proliferates.
The way IoT devices engage for collaborative purposes and implement security frameworks can be represented as graphs~\cite{george2018graph}.

\item \textbf{Epidemiology.}
The analysis of disease propagation and models of transition between states of health, infection, recovery and death are very important for public health and for ensuring standards of practices between countries to protect travelers and countries' populations~\cite{colizza2007predictability,bajardi2011human,brockmann2013hidden,chinazzi2020effect}.
These are represented as graphs, which can also be applied to localized health-related topics like reproductive health, sexual networks and the transmission of infections~\cite{liljeros2003sexual,bearman2004chains}.
They have even been used to model epidemics in massively multiplayer online games such as World of Warcraft~\cite{lofgren2007untapped}.
Real-life epidemics are perhaps at the forefront of examples of this application of graph theory for health preservation, with the most recent example as  COVID-19~\cite{surveillances2020epidemiological}.

\end{itemize}

\begin{figure*}[t]
	\centering
	\includegraphics[width=\textwidth]{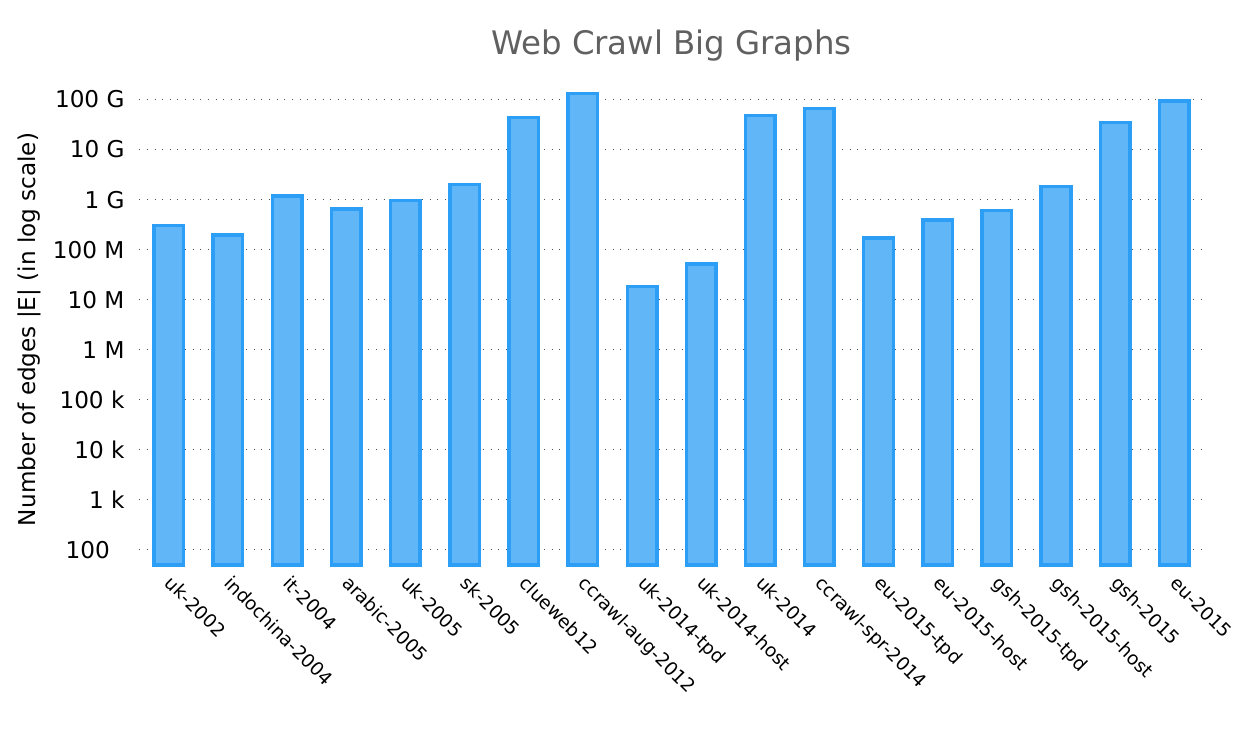}
	\caption{Web graph edge counts for domain crawls since the year 2000 (in log scale).}
	\label{fig:big-webgraph-sizes}
\end{figure*}

Other types of data represented as graphs can be found~\cite{sedgewick2011algorithms}.
To illustrate the growing magnitude of graphs, we focus on web graph sizes of different web domains in Fig~\ref{fig:big-webgraph-sizes}, where we show the number of edges for web crawl graph datasets made available by the \texttt{Laboratory} \texttt{of} \texttt{Web} \texttt{Algorithmics}~\cite{law_unimi_datasets} and by \texttt{Web} \texttt{Data} \texttt{Commons}~\cite{meusel2015graph}.
If one were to retrieve insights on the structure of these larger graphs (above a hundred million edges), it would become immediately clear that a combination of computer resources and specific software are necessary in order to process them.

\subsection{Motivation}\label{sec:introduction:sec:motivation}

We include this section in this survey to highlight three reasons.
Firstly, the recent years have seen a positive tendency in the field of all things related to graph processing.
As its aspects are further explored and optimized, with new paradigms proposed, there has been a proliferation of multiple surveys~\cite{Malicevic:2014:SGP:2592784.2592789,Han:2014:ECP:2732977.2732980,Kalavri:2014:ALG:2621934.2621940,kalavri2017high,heidari2018scalable,Sahu:2017:ULG:3186728.3164139,soudani2019investigation}.
They have made great contributions in systematizing the field of graph processing, by working towards a consensus of terminology and offering discussion on how to present or establish hierarchies of concepts inherent to the field.
Effectively, we have seen vast contributions capturing the maturity of different challenges of graph processing and the corresponding responses developed by academia and industry.

The value-proposition of this document is therefore, on a first level, the identification of the dimensions we observe to be relevant with respect to graph processing.
This is more complex than, for example, merely listing the types of graph processing system architectures or the types of communication and types of coordination within the class of distributed systems for graph processing.
Many of these dimensions, if not all, are interconnected in many ways.
As the study of each one is deepened, its individual overlap with the others is eventually noted.
For example, using distributed systems, it is necessary to distribute the graph across several machines.
This necessity raises the question of how to partition the graph to distribute it.
Afterward, as a natural consequence, it is necessary to define the coordination (e.g. synchronous versus asynchronous) between nodes of the system.
Orthogonally to this, the relation between data and computation in graph processing must be defined.
The vertices of a graph may represent posts in a social network, while the edges dictate follower relationships.
But to the systems that process them, one could specify the units of computation to be solely the vertices, exclusively the edges or components of the graph (and the definition of \textit{component} in this case would be required too).
Herein, we first note the cornerstone aspects of the graph processing field from individual works and from existing surveys: dimensions and definitions.

Secondly, we provide an exhaustive list of systems for graph processing, sharing their year of debut and distinctive features.
We explore the different types that exist:

\begin{itemize}
\item Single-machine: they may simply require copious amounts of memory, or instead employ compression techniques for graph processing.
\item Multi-machine: distributed systems which can be a cluster of machines (either homogeneous or heterogeneous) or special-purpose high-\-per\-formance computing systems (HPC).
\end{itemize}

On the third level, and also considering a scope of meta-analysis, we discuss the structuring of the field that is presented in existing surveys.
We complement it with our own highlight of important relations between graph processing concepts and a chronological analysis of the field of graph processing.

\subsection{Document Roadmap}\label{sec:introduction:sec:roadmap}

This following survey sections are as follows: 
\begin{itemize}
\item \textit{Background on Graph Notation and Terminology} covers the most essential notation used in graph theory.
\item \textit{Property Model and Query Languages} describes the systematizations developed to structure the execution of queries over graph data using computational resources.
\item \textit{Graph Algorithms: Natures and Types} highlights relevant aspects of graph-processing tasks and different types of graph algorithms.
\item \textit{Computational Representations} details important computational representations of graphs which typically use compression techniques.
\item \textit{Profile of Developers and Researchers} delves into results of a study on users of graph processing applications to identify important use-case trends.
\item \textit{Graph Processing: Computational Units and Models} analyzes how graphs are conceptually manipulated in many contributions of the field's state of the art, and the different levels of granularity.
\item \textit{Dimension: Partitioning} presents the most-known approaches to decomposing graph-based data into computational units for parallelism and distribution, showcasing models with different levels of granularity.
\item \textit{Dimension: Dynamism} enumerates scenarios with different definitions of \textit{dynamism} in graph processing, from the graph representing temporal data to the manipulated graph representation being constantly updated with new information.
\item \textit{Dimension: Workload} discusses the nature of graph processing workloads and different scopes such as analytics and storage.
\item \textit{Single-Machine and Shared-Memory Parallel Approaches} presents these types of architecture and describes important state-of-the-art marks.
\item \textit{High-Performance Computing} is focused on multi-core and multi-processor architectures.
\item \textit{Distributed Graph Processing Systems} enumerates systems which focus on distributing the graph across machines to enable processing.
\item \textit{Graph Databases} enumerates the graph database solutions we found as part of our research.
\item \textit{Conclusion} and final remarks.

\end{itemize}


\section[Background on Graph Notation and Terminology]{Background on Graph\\ Notation and Terminology}\label{sec:terminology}

Here we detail terms and concepts which are known in graph theory.
We include preliminary notions that serve as a basis to familiarize the reader with the language used in scientific documents on graph applications, processing systems and novel tech\-ni\-ques.
In the literature~\cite{Newman:2010:NI:1809753}, a graph $G$ is written as $G = (V, E)$ -- it is usually defined by a set of vertices $V$ of size $n = |V|$ and a set of edges $E$ of size $m = |E|$.
Vertices are sometimes referred to as \textit{nodes} and edges as \textit{links}.
An undirected graph $G$ is a pair $(V, E)$ of sets such that $E \subseteq V \times V$ is a set of unordered pairs.
If $E$ is a set of ordered pairs, then we say that $G$ is a directed graph.
Between the same two vertices there is usually at most one edge; if there are more, then the graph is called a \textit{multigraph} (note: an ordered graph in which a pair of vertices share two edges in opposite direction is not necessarily a multigraph).
Multigraphs are more common when looking at the applications and use-cases for graph databases such as \texttt{Neo4j}~\cite{miller2013graph}, where one may model more than one relation type between the same vertices.

Additionally, given a graph $G = (V, E)$, the set of vertices of $G$ is written as $V(G)$ and the set of edges as $E(G)$.
More commonly, we write $V(G) = V$ and $E(G) = E$.
The concept of a vertex's \textit{surrounding} is important for specifying traversals (relevant when considering graph query languages~\cite{Holzschuher:2013:PGQ:2457317.2457351}) and also defining scopes and units of computation in graph processing~\cite{Malewicz:2010:PSL:1807167.1807184,Roy:2013:XEG:2517349.2522740,Tian:2013:TLV:2732232.2732238}.
Two vertices $u, v \in V$ are adjacent or neighbours if $(u, v)$ is an edge - if $(u, v) \in E$.
Given a vertex $v \in V(G)$, its set of neighbours is denoted by $N_{G}(v)$, or succinctly by $N(G)$ when clearing $G$ from the context.
The set of edges with $v$ as a source or target is written as $E(v) = \{(u, u') \in E(G)\ |\ v = u$ or $v = u'\}$.
More generally, for two sets $X, Y \subseteq V$, we denote $E(X)$ as the set of edges with an end in $X$: $E(X) = \{(u, v) \in E(G)\ |\ u \in X$ or $v \in X\}$ and denote as $E(X, Y)$ the set of edges with an end in $X$ and an end in $Y$: $E(X, Y) = \{(u,v) \in G\ |$ either $u \in X$ and $v \in Y$, or $v \in X$ and $u \in Y\}$.
Furthermore, the degree of a vertex is used to indicate its number of neighbours.
For a graph $G = (V, E)$ and a vertex $v \in V$, the degree:- of $v$ may simply be commonly written as $d_{G}(v)$ or $d(v)$, with $d(v) = |N(v)|$.
If $G$ is a directed graph, it is found in the literature to be written as $d_{in}(v)$ and $d_{out}(v)$ the number of incoming and outgoing neighbours of $v$, respectively.

These notations are the basic building blocks for graph theory and a mandatory learning topic for those who come from backgrounds such as Mathematics, Computer Science and many other fields.
We note that across decades, the field of graph theory has matured and refined the notations and systematization of how we research and approach problems represented as graphs. 
These advancements have been accompanied with an even faster rhythm of technological progress. 
In tune, so has evol\-ved how these notations and representations are translated to actual actionable computational tasks and data.
The last decades have seen the introduction of query languages and models to represent graphs from the perspective of managing computational resources and solving graph-oriented problems.

\section[Property Model and Query Languages]{Property Model and Query\\ Languages}\label{sec:property_model_gql}

In this section we detail languages, models and subsequent standards for querying graph-based data.
The property graph model defines the organization of interrelated data as nodes (vertices), relationships (edges) and the properties of both types of elements~\cite{neo4j_graph_db}.
Instances of this model are usually found coupled to the design of graph databases and at the level of graph query languages~\cite{rodriguez2015gremlin,van2016pgql,angles2018property,green2018opencypher}.
Figure~\ref{fig:property-graph-model} shows a depiction of the properties that may be associated to vertices and edges in the context of the property graph model.

The ISO SQL Committee has accepted on September 2019 the Graph Query Language project proposal~\cite{gql_standard}, to enable SQL users to use property graph style queries on top of SQL tables.
This will promote the interoperability between SQL databases and graph databases and constitutes an important mark in the approximation between graph-structu\-red data and the databases that support it, an area that has been studied for decades~\cite{sheng1999graph,he2008graphs}.
The language standards engineering team of \texttt{Neo4j} identify the following cornerstones regarding the growing popularity of the property graph model~\cite{plantikowtowards}:

\begin{enumerate}

\item An intuitive model geared for application developers.

\item Ability of rapid prototyping with schema-optio\-nality.

\item Availability of native graph databases.

\item Declarative query languages focusing on ease of use with graph pattern matching and visual syntax, top-down linear statement order and language composability.

\end{enumerate}

\subsection{Graph Query Languages}\label{sec:property_model_gql:sec:gql}

The access to elements offered by graph databases is performed by means of specific query languages.
There are even projects focused on analytics which offer the ability to explore datasets using graph query languages without actually running a graph database, such as \texttt{GRA\-DO\-OP}~\cite{DBLP:journals/pvldb/JunghannsKTGPR18} and \texttt{GraphFrames}~\cite{mishra2019graphframes}. 
Languages such as \texttt{Cy\-pher} even support queries on gra\-phs which produce new graphs (for example to represent specific entities and relationships of the original graph) upon which to run further queries.
Here we list some of the most relevant graph query languages (and proposals in the scope of graph query languages), projects that structure the access to graph-based data as well as open-source constructs that form their basis.


\begin{itemize}

\item \texttt{Cy\-pher}.
An evolving graph query language\cite{francis2018cypher} which debuted with \texttt{Neo4j}'s entrance in the field of graph databases.
There have been efforts to adopt and develop the language in an open-source approach~\cite{green2018opencypher}. 
\texttt{Cy\-pher} has been an open and evolving language as part of the \texttt{openCypher} project~\cite{opencypher_repo}. 
The project has members involved in aspects such as synergy of engineering efforts with \texttt{A\-pa\-che} \texttt{Spark}~\cite{Zaharia:2010:SCC:1863103.1863113}, a language group and even interoperability features for systems that use \texttt{Grem\-lin}.
This graph query language heavily influenced the ISO pro\-ject for creating a standard graph query language\cite{opencypher_site} and has a syntax familiar to developers with knowledge of \texttt{SQL}.
\texttt{Cy\-pher} queries may be run on the following databases: \texttt{Neo4j}, \texttt{Graphflow}\cite{kankanamge2017graphflow}, \texttt{Re\-dis\-Gra\-ph}\cite{cailliau2019redisgraph} and \texttt{SAP} \texttt{Hana} \texttt{Gra\-ph}\cite{rudolf2013graph} and all databases where \texttt{Grem\-lin} is supported~\cite{opencypher_4_gremlin}.
This language is also used to express computation in \texttt{GRA\-DO\-OP}~\cite{DBLP:conf/grades/JunghannsKAPR17} and \texttt{Python Ruruki}~\cite{python_ruruki}.

\item \texttt{Grem\-lin}. 
A graph traversal machine and language, developed in scope of the \texttt{Apache} \texttt{Tin\-ker\-Pop} project\cite{tinkerpop_site}.
This project's development and growth was promoted by the now-defunct \texttt{Titan} graph database~\cite{titan} which was forked into the open-source \texttt{Janus\-Graph} data\-base\cite{janus} and the commercial \texttt{Da\-ta\-Stax} \texttt{Enter\-prise} \texttt{Gra\-ph} solution.
The traversal machine of \texttt{Grem\-lin} is defined as a set of three components~\cite{rodriguez2015gremlin}: the data represented by a graph $G$, a traversal $\Psi$ (instructions) which consists of a tree of functions called \textit{steps}; a set of traversers $T$ (read/write heads).
With this composition, \texttt{Grem\-lin} and its traversal machine enable the exploration of multi-dimensional structures that \textit{model a heterogeneous set of ``things'' related to each other in a heterogeneous set of ways} as detailed in~\cite{rodriguez2015gremlin,DBLP:journals/corr/Rodriguez15}.
A traversal evaluated against a graph may generate billions of traversers, even on small graphs, due to the exponential growth of the number of paths that exist with each step the traversers take.
Databases supporting \texttt{Grem\-lin} include \texttt{Orient\-DB}\cite{tesoriero2013getting}, \texttt{Neo4j}\cite{Webber:2012:PIN:2384716.2384777}, \texttt{Da\-ta\-Stax} \texttt{Enterprise}\cite{datastax}, \texttt{In\-fi\-ni\-te\-Graph}\cite{objectivitydb}, \texttt{Janus\-Gra\-ph}\cite{janus}, \texttt{Azure} \texttt{Cosmos} \texttt{DB}\cite{paz2018introduction} and \texttt{Amazon} \texttt{Nep\-tu\-ne}\cite{bebee2018amazon}, while the graph processing systems that allow its use are \texttt{A\-pa\-che} \texttt{Gi\-ra\-ph}~\cite{ching2013scaling} and \texttt{A\-pa\-che} \texttt{Spark}~\cite{Armbrust:2015:SSR:2723372.2742797}.


\item \texttt{SPAR\-QL}. 
The standard query language for \texttt{RDF} data (triplets), also known as the query language for the semantic web.
We mention it due to its graph pattern matching capability~\cite{perez2009semantics} and scalability potential of querying large \texttt{RDF} graphs~\cite{huang2011scalable}.
As \texttt{RDF} is a direct labeled graph data format, \texttt{SPAR\-QL} becomes a language for graph-matching.
Its queries have three components: a \textit{pattern matching part} allowing for pattern unions, nesting, filtering values of ma\-tchings and choosing the data source to match by a pattern; a \textit{solution modifier} to allow modifications to the computed output of the pattern, such as applying operators as projections, orderings, limits and distinct; the \textit{output}, which can be binary answers, selections of values for variables that matched the patterns, construction of new \texttt{RDF} data from the values or descriptions of resources.
While graph databases may not necessarily be triplet stores, the graph query languages they support may allow for example that the \texttt{RDF}-specific semantics of a \texttt{SPAR\-QL} query may be translated to \texttt{Cy\-pher}, \texttt{Grem\-lin} or another language.
\texttt{SPAR\-QL} is also supported (analytics) over \texttt{GraphX}~\cite{schatzle2015s2x} and the higher-level graph analytics tool \texttt{Graph\-Frames}\cite{bahrami2017efficient}.
Among the graph databases that support this language we have \texttt{Amazon} \texttt{Nep\-tu\-ne}\cite{bebee2018amazon} and \texttt{Alle\-gro\-Gra\-ph}\cite{allegrograph} (the later two more oriented to the purpose of \texttt{RDF}).

\item \texttt{Gra\-ph\-QL}.
A framework developed and internally used at Facebook for years before its reference implementation was released as open-source~\cite{fb_graphql}.
It introduced a new type of web-based interface for data access.
As a framework, one of its core components is a query language for expressing data retrieval requests sent to web servers that are \texttt{Gra\-ph\-QL}-aware.
The queries are syntactically similar to \texttt{Java\-Script} \texttt{Object} \texttt{No\-ta\-tion} \texttt{(JSON)}.
The \texttt{Gra\-ph\-QL} specification implicitly assumes a logical data model implemented as a virtual, graph-based view over an underlying database management system~\cite{hartig2017initial}.
It has been studied with the semantics of its queries formalized as a labeled-graph data model and the total size of a \texttt{Gra\-ph\-QL} response shown to be computable in polynomial time~\cite{hartig2018semantics}.
\texttt{Gra\-ph\-QL} is more than a query language - it defines a contract between the back-end and front-end over an agreed-upon type system, forming an application data mo\-del as a graph.
It is useful as it proposes a decoupling between the back-end and front-end, allowing each component to be changed independently of the other.
For example, to serve queries in the graph, the data in the back-end could come from databases (e.g. \texttt{Neo4j} as the back-end serving \texttt{Gra\-ph\-QL} queries received at a web endpoint~\cite{neo4j_graphql}), in-memory representations or other APIs.

\item \texttt{PGQL}.
The \texttt{Property Graph Query Language}, based on the paradigm of graph pattern matching~\cite{van2016pgql}.
It closely follows the syntactic structures of \texttt{SQL}, providing regular path queries with conditions on labels and properties to enable reachability and path finding queries. 
The data types it defines are the intrinsic \textit{vertex}, \textit{edge}, \textit{path} and also an intrinsic \textit{graph} type, allowing for graph construction and query composition.
It was motivated by the fact that \texttt{SPAR\-QL} is the \texttt{RDF} standard query language, thus imposing that graphs be represented as a set of triples (or edges), and by \texttt{Cy\-pher}'s lack of support for regular path queries and graph construction as fundamental graph querying functionalities.
\texttt{PGQL} also provides tabular output, allowing its queries to be naturally nested inside \texttt{SQL} queries, allowing for easy integration into existing database technology.
It is used in Oracle's products. 

\item \texttt{G-Log}. 
A declarative query language on graphs which was designed to combine the expressiveness of logic, the modeling of complex objects with identity and the representations enabled by graphs~\cite{paredaens1995g}.
The authors describe it as a \textit{deductive language for complex objects with identity}, with a data model that captures the modeling capabilities of object-oriented languages, although lacking their typical data abstraction features which are related to system dynamism.
They claim \texttt{G-Log} may be seen as a graph-based symbolism for first-order logic and they prove that all sentences of first-order logic may be written in \texttt{G-Log}. 
Secondly, they define the semantics of the language for data\-ba\-se query evaluation.
Lastly, the authors sta\-te that due to being \textit{a very powerful language}, its computation could be unnecessarily inefficient in the most general case.
We mention \texttt{G-Log} as it is historically relevant due to highlighting the importance and expressiveness of graph-based data models in manipulating the relationships between data.

\end{itemize}

A study on these aspects inherent to graph-stru\-ctu\-red data has been performed using different data models (\texttt{RDF}, property graph and relational model) and a sample of systems (\texttt{RDF}: \texttt{Vir\-tu\-oso}~\cite{erling2012virtuoso}, \texttt{Tri\-ple\-Ru\-sh}\cite{stutz2013triplerush}; gra\-ph data\-bases: \texttt{Neo\-4j}, \texttt{Spark\-see}\-\cite{martinez2011dex}; re\-la\-tional database: \texttt{Vir\-tu\-oso}) that offer the models ~\cite{gubichev2014graph}.
These systems were compared using \texttt{LUBM}, a well-known and used synthetic \texttt{RDF} bench\-mark\cite{guo2005lubm}.
The authors make interesting observations: the verbosity of implementing queries in \texttt{Spark\-see} due to its design of delegating the execution plan implementation to the developer, as opposed to using declarative languages like \texttt{Cy\-pher}; the importance of cost-based query optimization wi\-thin query engines.
Query patterns used in the literature fall in the following categories~\cite{van2016pgql,gubichev2014graph}: single triple patterns satisfying a given condition; matches on the nodes that are adjacent and edges that are incident to a given node; triangles, which look for three nodes adjacent to each other; paths of fixed or variable length.

We see in Table~\ref{table:gql_databases} that \texttt{Cy\-pher} and \texttt{Grem\-lin} have seen broad use in both storage and analytics tasks, with the latter also used in \texttt{RDF} databases.
\texttt{SPAR\-QL} is mostly supported by \texttt{RDF} databases and some frameworks and libraries have interoperability with it.
The \texttt{Gra\-ph\-QL} compatibility row in Table~\ref{table:gql_databases} signals that the systems with check marks have clients for its structure. 
\texttt{Gra\-ph\-QL}'s decoupling of front-end and back-end has support by \texttt{GraphDB} and also \texttt{Neo4j Cypher}~\cite{neo4j_graphql_cypher} and \texttt{OrientDB Gremlin}.
We did not present an exhaustive list of the \texttt{RDF}-specific graph databases; only the most relevant with relationships to most graph query languages.


There have been studies comparing the performance of graph query languages between themselves and native access.
Comparisons have been made between \texttt{Cy\-pher}, \texttt{Grem\-lin} and native access regarding the ease in expressing queries and the performance gains in \texttt{Neo4j}~\cite{Holzschuher:2013:PGQ:2457317.2457351}.
The authors subjectively note that \texttt{Cy\-pher} will \textit{feel more natural} for developers with existing experience with \texttt{SQL} languages and could be considered more high-level than \texttt{Grem\-lin}.
They further observe that \texttt{Grem\-lin} queries achieved better performance than with \texttt{Cy\-pher} in some query types.
There have been other studies on the properties of graph query languages across the decades~\cite{amann1993gram,haase2004comparison,wood2012query,angles2017foundations,angles2018g}, which emphasizes the relevance of the theoretical and pragmatic aspects of graph query language development, whose renewed interest by industries is proof of it being an important part of extracting value from data.
We end this section on graph query languages by again drawing attention to the recent global coordination to create a standard graph query language - it draws from the lessons of different graph query languages and involves the collaboration of many academic and industrial partners whose individual experiences are helping to shape the future of the way graphs are queried. 

The high-level solutions and languages we describe are able to translate into efficient computations. 
They bring the benefit of standardizing and approaching theoretical representations to query spe\-ci\-fi\-cation.
However, such translation is not direct, as the nature of the computation to be performed on the graph and the type of algorithm to execute have a profound impact on how computational resources should be optimized.

\section[Graph Algorithms: Natures and Types]{Graph Algorithms: Natures\\ and Types}\label{sec:graph_algorithms}

There are several aspects inherent to gra\-ph-pro\-ce\-ssing tasks.
Graphs have properties which may be extrapolated using specific algorithms, from computing the most important vertices (e.g. using an arbitrary function like PageRank~\cite{pageRank}), finding the biggest communities (for which there is a choice of many algorithms) or the most \textit{relevant} paths (for a definition of relevancy).
An algorithm that processes the whole graph (as opposed to localized information queries expressed with graph query languages seen previously) is typically executed in parallel fashion when the resources for parallelism are available.
When implementing these algorithms, whether the developer manually implements the parallelism or merely uses such a functionality offered by an underlying framework (e.g. \texttt{Apache} \texttt{Spark}\cite{Gonzalez:2014:GGP:2685048.2685096} or \texttt{Apache} \texttt{Flink}\cite{carbone2015apache}), some challenges must be considered.
This means that while the field of graph processing is developed with the goal of improving how we manipulate and extract value from graph-based data, as the techniques to achieve this end become more refined, other aspects of graph structures gain prominence as challenges to them.

We list and comment inline here the major types of challenges of parallel graph processing identified in a previous study~\cite{lumsdaine2007challenges}:

\begin{enumerate}

\item \textit{Data-driven computations}: a graph has vertices and edges which establish how computations are performed by algorithms, making graph applications data-driven.
We see this observation shift the focus to data - what shou\-ld the elementary unit of computation be?
In this survey we go over multiple solutions in the literature, considering the computation from the perspective of vertices~\cite{Malewicz:2010:PSL:1807167.1807184,DBLP:books/sp/SOAK2016,carbone2015apache,Gonzalez:2014:GGP:2685048.2685096}, edges~\cite{Roy:2013:XEG:2517349.2522740} and sub-graphs~\cite{Tian:2013:TLV:2732232.2732238}. 

\item \textit{Irregular problems}: the distribution of edges and vertices usually does not constitute uniform graphs that form embarrassingly parallel problems, whose benefits from simple parallelism are easier to achieve~\cite{wilkinson2004parallel}.
We note that after defining the unit of computation in a graph, care needs to be taken when assigning parts of the graph to different processing units.
Skewed data will negatively impact load balance~\cite{Kalavri:2014:ALG:2621934.2621940} unless tailored approaches are undertaken, which take into account different types of graph properties such as scale-free~\cite{fortunato2006scale} in their designs~\cite{Gonzalez:2012:PDG:2387880.2387883,chen2015powerlyra}.

\item \textit{Poor locality}: locality-based optimizations offered by many processors are hard to apply to the inherently irregular characteristics of graphs due to poor locality during computation.
We believe it is important to mention how this manifests when using distributed systems (clusters) to process the graphs.
To mitigate this, techniques may be used for example to replicate specific vertices based on properties such as their degree, or to use specific graph partitioning strategies when working wi\-th vertex-centric approaches~\cite{soudani2019investigation}.

\item \textit{High data-access-to-computation ratio}: the authors note that a large portion of graph processing is usually dedicated to data access in graph algorithms and so waiting for memory or disk fetches is the time-most consuming phase relative to the actual computation on a vertex or edge itself.
We note one approach~\cite{Roy:2015:CSG:2815400.2815408} to this problem that focused on balancing network and storage latencies with computation time to minimize the impact of underlying data accesses in a cloud computing setting.

\end{enumerate}

\subsection{Algorithms}\label{sec:sec:graph_algo}

Graph algorithms that execute globally over all elements of a graph have a distinct nature from those solved with graph query languages - the scope of computation is drastically different with respect to the computational resources needed to satisfy it. 
We note, however, some graph databases such as \texttt{Neo4j} have extensions like the \texttt{Neo4j APOC Library} for languages like \texttt{Cy\-pher} to start algorithms with global computation from the graph que\-ry lan\-gua\-ge\cite{neo4j_apoc_lib}.
A previous survey on the scalability of graph processing frameworks~\cite[Sec. 3.3]{dominguez2010discussion} defines a categorization of graph algorithms, which we reproduce here:


\begin{itemize}

\item \textit{Traversals}.
Starting from a single node, they employ recursive exploration of the neighbourhood until a termination criteria is met, like reaching a desired node or a certain depth.
Instances of this are for example calculating single-source shortest-paths (SSSP), $k$-hop nei\-ghborhood or breadth-first searches (BFS).

\item \textit{Graph analysis}.
Algorithms falling into this scope aim at understanding the structural properties and topology of the graph.
They may be executed to grasp properties like the diameter (greatest distance between any pair of vertices), density (ratio of the number of edges $|E|$ with respect to the maximum possible edges) or degree distribution.

\item \textit{Components}.
Concept: a \textit{connected component} is a subset of graph vertices for which there is a path between any pair of vertices.
Finding connected components is relevant to detect frailties in the networks that graphs represent.
The connections between these components are called \textit{bridges}, which if removed, will separate connected components.

\item \textit{Communities}.
The groups called \textit{communities} consist of sets of vertices that are close to each other within a community than to those outside it.
There are different techniques to compute them such as minimal-cut sets and label propagation, among others.

\item \textit{Centrality measures}.
These represent the importance of a vertex with respect to the rest of the network.
Many definitions exist such as PageRank and betweenness centrality, for example.
There also heuristics to measure the relevance of edges such as spanning edge betweenness~\cite{teixeira2013spanning}.

\item \textit{Pattern matching}.
Related to algorithms aimed at recognizing or describing patterns, known as \textit{graph matching} in this context.

\item \textit{Graph anonymization}.
To produce a graph with similar structure but making it so the entities represented by vertices and edges are not identifiable.
Two examples of anonymization procedures are $k$-degree and $k$-neighbourhood anonymity.

\end{itemize}

%


%

Underlying the computations that take place to solve graph-specific tasks lies the granularity. 
How are the vertices, edges and properties of graphs processed or stored? 
This required building a bridge from these mathematically-defined elements to the bits and bytes of computers.

\newpage

\section[Computational Representations]{Computational Representa-\\tions}\label{sec:computational_representations}

Underlying all these ways to extract information from graphs is their digital representation.
It is important to understand the set of operations to be performed over the graph and its size in order to guide the choice of representation.
To represent the edges, perhaps the two most well-known approaches are the adjacency list and adjacency matrix.
The choice of using an adjacency list or a matrix usually depends on the amount of edges in the graph.

\begin{figure*}
	\centering
	\subfigure[Sample graph $G$]
	{
		\begin{tikzpicture}[show background rectangle,->,>=stealth',
	shorten >=1pt,auto,node distance=1.25cm,
    semithick,main node/.style={circle,draw}]

	\node[main node] (1) {1};
	\node[main node] (2) [left of=1] {2};
	\node[main node] (3) [below of=1] {3};
	\node[main node] (4) [below of=2] {4};
	\node[main node] (5) [right of=3] {5};
	\node[main node] (6) [right of=1] {6};

	\path
		(1) edge [bend right] node {} (2)
		(2) edge node [below] {} (1)
			  edge node [below] {} (4)
		(3) 
			edge node[right] {} (1)
			edge node[below] {} (2)
		(4) edge node [below] {} (3)
		edge node [below] {} (6)
		(6) edge node [below] {} (5)
		edge node [left] {} (1);
	\draw [->] [out=45,in=-45] (5) to (6);

\end{tikzpicture}
		\label{fig:graph_representations:graph}
	}
	\subfigure[Adjacency matrix $A$]
	{
		\begin{tikzpicture}
	\matrix[ampersand replacement=\&,matrix of math nodes,
			left delimiter={[},
			right delimiter={]},
			nodes in empty cells] (m)
	{
		0 \& 1 \& 0 \& 0 \& 0 \& 0 \\
		1 \& 0 \& 0 \& 1 \& 0 \& 0 \\
		1 \& 1 \& 0 \& 0 \& 0 \& 0 \\
		0 \& 0 \& 1 \& 0 \& 0 \& 1 \\
		0 \& 0 \& 0 \& 0 \& 0 \& 1 \\
		1 \& 0 \& 0 \& 0 \& 1 \& 0 \\
	};
\end{tikzpicture}
		\label{fig:graph_representations:adj-matrix}
	}
	\subfigure[Adjacency list]
	{
		\input{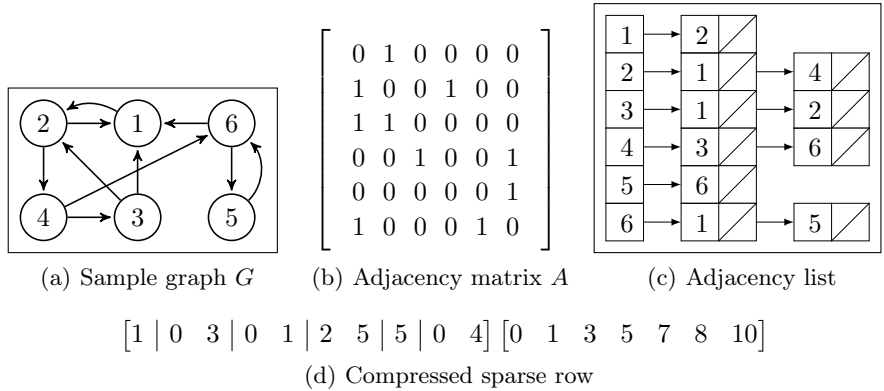}
		\label{fig:graph_representations:adj-list}
	}
	\subfigure[Compressed sparse row]
	{
		\begin{tikzpicture}
	\node (b) [inner sep=0pt] {
    $\begin{bmatrix}[c|cc|cc|cc|c|cc]
       1 & 0 & 3 & 0 & 1 & 2 & 5 & 5 & 0 & 4
    \end{bmatrix}
	\begin{bmatrix}
	   0 & 1 & 3 & 5 & 7 & 8 & 10
    \end{bmatrix}$
	};
\end{tikzpicture}
		\label{fig:graph_representations:adj-csr}
	}
	\caption{Simple computational representations of sample directed graph $G$ shown in (a).}	
	\label{fig:graph_representations}
\end{figure*}

Consider a given graph $G = (V, E)$.
If $|E|$ is close to the maximum number of edges that a graph can sustain ($|E| \simeq |V|^2$), then it is a \textit{dense} graph and it makes more sense to choose the adjacency matrix (performance-wise).
However, if the graph is \textit{sparse} ($|E| \ll |V|^2$), where most nodes are not connected, it can be efficiently represented (storage-wise) with an adjacency list.
While the matrix consumes more space than the adjacency list, it allow for constant-time access.
We show in Fig.~\ref{fig:graph_representations} an example of different representations for the same graph.

Figure~\ref{fig:graph_representations:graph} shows a sample graph $G$, for which the adjacency matrix is shown in Fig.~\ref{fig:graph_representations:adj-matrix} and the corresponding adjacency list in Fig.~\ref{fig:graph_representations:adj-list}.
The first row of the adjacency matrix $A$ represents the outgoing edges of vertex $1$, which is connected to vertex $2$.
It is common in the literature~\cite[Ch. 22]{10.5555/1614191} to use the subscript notation $A_{i,j}$ to refer to the presence of a specific edge in matrix $A$ (the notation is relevant for theoretical purposes even if using another type of representation) starting from vertex $i$ and targeting vertex $j$:
\begin{equation}
	A_{i,j}  =
	\begin{cases}
		1\ \text{if there is an edge from}\ i\ \text{to}\ j,\\    
		0\ \text{otherwise.}
	\end{cases}
\end{equation}
Matrix $A$ also takes on a particular configuration depending on the graph being directed or undirected.
In the later case, there is no explicit sense of source or target of an edge, leading to symmetry in matrix $A$.
Implementations of graph processing systems often represent undirected graphs as directed graphs such that the undirected edge between a pair of vertices is represented by two directed edges in opposite directions between the pair.

There are more space-efficient ways to represent a graph, and they become a necessity when exploring the realm of big graphs.
The choice between an adjacency list or matrix is bound to the density of the graph.
But to justify other representation types, factors such as graph size, storage limitations and performance requirements need special focus.
The compressed sparse row (CSR), also known as the \textit{Yale format}, is able to represent a matrix 
using three arrays: one containing nonzero values; the second containing the extents of rows; the third storing column indices.
Figure~\ref{fig:graph_representations:adj-csr} shows a representation in this format (we omit the array the array containing the nonzero values as they are all one in this case).

Let us consider that indices are zero-based. 
The array on the left side is the column index array, where the pipe character $|$ separates groups which belong to each row of $A$. 
The second row (index 1) of $A$ has elements at position indexes 0 and 3 in $A$. 
Therefore, the second group of the column index array has elements $[0, 3]$.
The array on the right is the row index array which has one element per row in matrix $A$ and an element which is the count of nonzero elements of $A$ at the end of the array (there are variations without this count).
For a given row $i$, it encodes the start index of the row group in the column index array (on the left in Fig.~\ref{fig:graph_representations:adj-csr}).
This way, for example, the second row of matrix $A$ (Fig.~\ref{fig:graph_representations:adj-matrix}) has row index 1 in $A$. 
Then, looking at the row index array (the one on the right), as the second row of matrix $A$ has row index 1, we access the elements with indices $[1, 2]$ in the row index array, which returns the pair $(1, 3)$, indicating that the second row (index one) of $A$ is represented in the column index array starting (inclusive) at index 1 and ending at index 3 (exclusive).
If we look at the column index array and check the elements from index 1 (inclusive) to 3 (exclusive), we get the set of values $\{0, 3\}$.
And if we look at the second row in $A$, column index 0 and column index 3 are exactly the positions of the edges in $A$ for that row.
Generally, for a matrix $M$'s row index $i$, we access indices $[i, i+1]$ in the row index array, and the returned pair dictates the starting (inclusive) and ending (exclusive) index interval in the column index array. 
The set of elements in that interval in the column index array contains the indices of the columns with value 1 for row index $i$ in $M$.
We point the reader to~\cite{bulucc2009parallel} for details on its representation and construction.
There is also the compressed sparse column (CSC), which is similar but focused on the columns, as the name suggest.

Other approaches take advantage of domain-spe\-ci\-fic properties of graphs. 
Such is the case of \texttt{Web\-Graph}\cite{DBLP:conf/www/BoldiV04}, which exploits certain properties of web graphs to represent them with increased compression.
An important property they exploit is \textit{locality}, as many links stay within the same domain, that is, if the web graph is lexicographically ordered, most links point close by.
Another property is \textit{similarity}: pages that are close by in the lexicographical order are likely to have sets of neighbours that are similar.
The study performed with \texttt{Web\-Graph} also highlighted, among other facts, the following: similarity was found to be much more concentrated than previously thought; consecutivity is common regarding web graphs. 
The properties of ordering (and different techniques to produce them) have also been exploited by the same authors to obtain compression with social networks.
\texttt{Web\-Graph} was used in an extensive analysis of many different data sets, which were made available online by the \texttt{Laboratory for Web Algorithmics}~\cite{law_unimi_datasets,DBLP:conf/www/BoldiV04,Boldi:2011:LLP:1963405.1963488,BCSU3,BMSB}.

\begin{figure*}
	\centering
	\subfigure[Decomposition ($k = 2$)]
	{
	\resizebox{0.35\textwidth}{!}{
		\begin{tikzpicture}
	\matrix[ampersand replacement=\&,matrix of math nodes,
			row sep=0.40,column sep=0.40,
			left delimiter={[},
			right delimiter={]},
			nodes in empty cells] (m)
	{
		1 \& 0 \& 0 \& 0 \& 0 \& 0 \& 0 \& 0 \\
		0 \& 0 \& 0 \& 0 \& 0 \& 0 \& 0 \& 0 \\
		0 \& 0 \& 1 \& 0 \& 0 \& 0 \& 0 \& 0 \\
		0 \& 0 \& 0 \& 0 \& 0 \& 0 \& 0 \& 0 \\
		0 \& 0 \& 0 \& 0 \& 1 \& 0 \& 0 \& 0 \\
		0 \& 0 \& 0 \& 0 \& 0 \& 0 \& 0 \& 1 \\
		0 \& 0 \& 0 \& 0 \& 0 \& 0 \& 0 \& 0 \\
		0 \& 0 \& 0 \& 0 \& 0 \& 0 \& 0 \& 0 \\
	};
	\draw (m-1-1.north west) rectangle (m-4-4.south east-|m-1-4.north east);
		\draw (m-1-1.north west) rectangle (m-2-2.south east-|m-1-2.north east);
			\draw (m-1-1.north west) rectangle (m-1-2.south east-|m-1-1.north east);
			\draw (m-2-2.north west) rectangle (m-2-3.south east-|m-2-2.north east);
		\draw (m-3-3.north west) rectangle (m-4-4.south east-|m-3-4.north east);
			\draw (m-3-3.north west) rectangle (m-3-4.south east-|m-3-3.north east);
			\draw (m-4-4.north west) rectangle (m-4-5.south east-|m-4-4.north east);
	\draw (m-5-1.north west) rectangle (m-8-4.south east-|m-5-4.north east);
	\draw (m-1-5.north west) rectangle (m-4-8.south east-|m-1-8.north east);
	\draw (m-5-5.north west) rectangle (m-8-8.south east-|m-5-8.north east);
		\draw (m-5-5.north west) rectangle (m-6-6.south east-|m-5-6.north east);
			\draw (m-5-5.north west) rectangle (m-5-6.south east-|m-5-5.north east);
			\draw (m-6-6.north west) rectangle (m-6-7.south east-|m-6-6.north east);
			
			\draw (m-5-7.north west) rectangle (m-5-8.south east-|m-5-7.north east);
			\draw (m-6-8.north west) rectangle (m-6-8.south east-|m-6-8.north east);
		\draw (m-7-7.north west) rectangle (m-8-8.south east-|m-7-8.north east);
\end{tikzpicture}
		}
	\label{fig:k2-tree-construction:adj-k2-matrix}
	}
	\subfigure[$k^2$-tree ($h = \lceil\log_{k^2}{n}\rceil = \lceil\log_{4}{64}\rceil = 3$)]
	{
	\resizebox{0.555\textwidth}{!}{
		\input{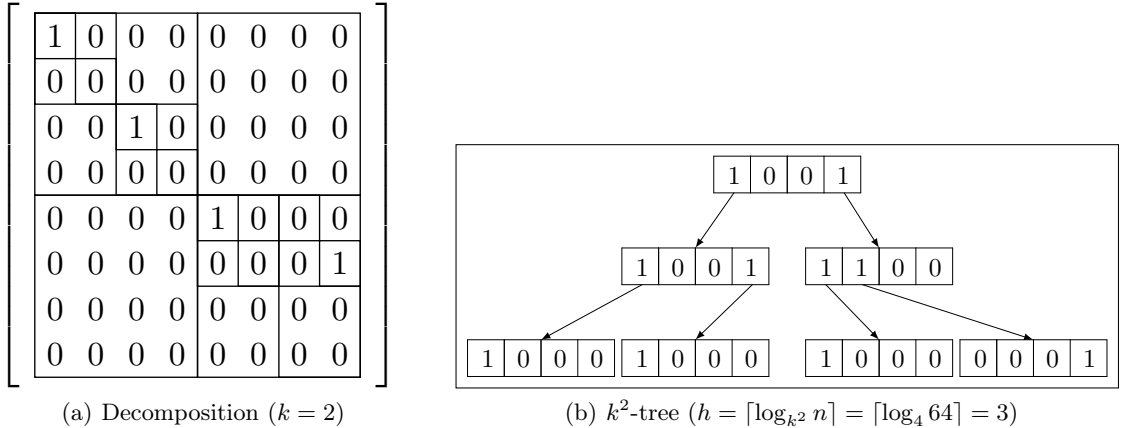}
		}
	\label{fig:k2-tree-construction:adj-k2-tree}
	}
	\caption{A sample adjacency matrix ($n = |V| = 64$) and corresponding $k^2$-tree representation.}
	\label{fig:k2-tree-construction}
\end{figure*}

The $k^2$-tree is another data structure employed to represent and efficiently store graphs~\cite{Brisaboa2014}.
It may be used to represent static graphs and binary relations in general.
It has been used to represent binary relations like web graphs, social networks and \texttt{RDF} data sets by internally using compressed bit vectors.
Conceptually, we recursively subdivide each block of a graph's adjacency matrix until we reach the level of individual cells of the matrix.
The idea is to divide (following an \texttt{MX-Quadtree} strategy~\cite[Sec. 1.4.2.1]{samet2006foundations}) the matrix in blocks and then assign 0 to the block if it only contains zeros (no edges) or 1 if it contains at least an edge.
We show in Fig.~\ref{fig:k2-tree-construction} a sample adjacency matrix on the left and the corresponding $k^2$-tree representation of the decomposition.
This representation of the adjacency matrix is actually a $k^2$-tree of height $h = \lceil\log_{k}{n}\rceil$, where ($n = |V|$ and) each node contains a single bit of data.
It is 1 for internal nodes and 0 for leaves, except for the last level, in which all nodes are leaves representing values from the adjacency matrix.
It is a data structure that also efficiently matches the properties of sparseness and clustering of web graphs.

Another proposal, \texttt{Log(Graph)}~\cite{besta2018log} is a graph representation that combines high compression ratios with low overhead to enable competitive processing performance while making use of compression. 
It achieved compression ratios similar to \texttt{Web\-Graph} while reaching speedups of more than 2x.
The authors achieve results by applying logarithm-based approaches to different graph elements.
They describe its application on \textit{fine elements} of the adjacency array (the basis of \texttt{Log(Graph)}: vertex IDs, offsets and edge weights.
From information theory, the authors note that a simple storage lower bound can be the number of possible instances of an entity, meaning the number of bits required to distinguish them.
Using this type of awareness on the different elements that represent an adjacency array and by incorporating bit vectors, the authors present a \texttt{C++} library for the development, analysis and comparison of graph representations composed of the many schemes described in their work.



There were relevant techniques for graph compression the literature on graph compression~\cite{feder1995clique,apostolico2009graph,buehrer2008scalable,kang2011beyond,fan2012query,lim2014slashburn} with the \texttt{Web\-Graph} framework~\cite{DBLP:conf/www/BoldiV04,DBLP:conf/dcc/BoldiV04} as one of the most well-known, and more recently the $k^2$-tree structure~\cite{Brisaboa2014,DBLP:journals/kais/HernandezN14,DBLP:conf/dcc/BrisaboaBGLPT15,gagie2015faster}, only later was the focus cast on being able to represent big graphs with compression while allowing for updates.
Furthermore, if we add the possibility of dynamism of the data (the graph is no longer a static object that one wishes to analyze) to the factors guiding representation choice, then it makes sense to think about how to represent a big graph in common hardware not only for storage purposes but also for efficient access with mutability. 
Works such as \texttt{VeilGraph}~\cite{coimbra2019veilgraph} approach the concepts of efficient representations by for example incorporating summary graph structures to reduce the total performed computations in the context of graph updates.

A dynamic version of the $k^2$-tree structure was proposed for this purpose~\cite{DBLP:journals/is/BrisaboaCBN17}.
Using compact representations of dynamic bit vectors to implement this data structure, the $k^2$-tree was used to provide a compact representation for dynamic graphs.
However, this representation with dynamic compact bit vectors suffers from a known bottleneck in compressed dynamic indexing~\cite{navarro2016compact}.
It suffers a logarithmic slowdown from adopting dynamic bit vectors.
A recent comparative study on the graph operations supported by different $k^2$-tree implementations has also been performed~\cite{MiguelE.Coimbra_1_2020}.
This work also presented an innovative take on implementing dynamic graphs by employing the $k^2$-tree data structure with a document collection dynamization technique~\cite{DBLP:conf/pods/MunroNV15}, avoiding the bottleneck in compressed dynamic indexing.

To better understand the dynamics of these representations, we next analyze recent work on profiling the needs of practitioners of graph processing.

\section{Profile of Developers and Researchers}\label{sec:devs_researchers}

There has been a recent study on the profile of practitioners in both academia and industry~\cite{Sahu:2017:ULG:3186728.3164139}.
To our knowledge, it has been the first one to identify the profile of users and the types of computational tasks they need to process over graphs.
Specifically, this is the result of an online survey aiming to establish: $1)$ the types of graph data used; $2)$ the computations users run on their graphs; $3)$ the softwares used to perform computations; $4)$ the major challenges faced by users when processing the graph data.
The authors share five main findings as a result of their online survey:

\begin{itemize}

\item There is \textbf{variety in the data} represented with graphs.
From participants, they found that enterprise data such as orders, transactions and products are a very common form of data represented as graphs, when they initially would seem to be the type of data that is perfect for relational database systems.

\item Very large graphs are present from \textbf{small to large enterprises}, which contradicts the notion that very large graphs are only present in the challenges of large corporations like Amazon, Google.
We can therefore claim that there is still immense value regarding innovative solutions that allow the processing of big graphs under the potentially-scarce resources held by the smaller companies and startups.

\item As a consequence, the \textbf{ability to process ve\-ry large graphs efficiently} is still the biggest limitation of existing software.
We see in this the relevance of novel techniques underlying the design of graph processing systems (both distributed and not) to reduce overhead factors such as storage and network latency, coordination (between threads in CPUs and between nodes in a cluster) and redundancy (for example consider the optimization of replication factor in vertex-centric systems).

\item The second most desired ability is \textbf{visualization}, related to challenges in graph query languages.
Visualizing all vertices in a graph qui\-ckly becomes a prohibitive endeavor as graph size increases.
However, being able to efficiently visualize structural aspects of the graph (e.g. communities) or relevant nodes under some criteria (e.g PageRank~\cite{pageRank}) is important.
The\-se same capabilities are also relevant for graph query languages.
When coupled with graph databases (e.g. \texttt{Neo4j}), there may be a visualization layer to give direct feedback of the queries that are being written.
We note this potential in visualization gains another dimension if the graph is changing - it may be more efficient to update existing visualizations with approximations based on graph update deltas as opposed to recalculating the visualizations from the beginning on the updated graph.

\item Relational database management systems \textbf{(R\-D\-BMSes) are still relevant} in managing and processing graphs.

\end{itemize}

This survey~\cite{Sahu:2017:ULG:3186728.3164139} was important as it highlighted that graphs falling in the realm of big data are still the top priority for innovative solutions, thus further motivating the importance of concisely defining the landscape of graph processing and different ways in which the need to process big graphs manifests.
The processing of a graph can take place from different perspectives such as vertices, edges or parts of the graph. 
These perspectives can more naturally represent one type of graph computation than other - herein we analyze them.

\section[Graph Processing: Computational Units and Models]{Graph Processing: Computa-\\tional Units and Models}\label{sec:computational-units}

Here we detail the most relevant paradigms and computational units used to express computation in graph processing systems.
Programming models for gra\-ph processing have been studied and documented in the literature~\cite{kalavri2017high, heidari2018scalable}.
They define properties such as the granularity of the unit of computation, how to distribute it across the cluster and how communication is performed to synchronize computational state across machines.

\subsection{Unit: Vertex-Centric (TLAV)}\label{sec:computational-units:sec:vertex-centric}

The vertex-centric paradigm, also known as \textit{think-like-a-vertex} (TLAV), debuted with Google's \texttt{Pre\-gel} system~\cite{Malewicz:2010:PSL:1807167.1807184}.
An open-source implementation of this model known as \texttt{A\-pa\-che} \texttt{Gi\-ra\-ph}~\cite{ching2013scaling} was then offered to the public.
Other example systems that were created using that model are \texttt{Gra\-ph\-Lab}~\cite{DBLP:journals/corr/LowGKBGH14}, \texttt{PowerGraph}~\cite{Gonzalez:2012:PDG:2387880.2387883}, \texttt{Pow\-er\-Ly\-ra}~\cite{chen2015powerlyra}.
As the unit of computation is the vertex itself, the user algorithm logic is expressed from the perspective of vertices.
The idea is that a vertex-local function will receive information from the vertex's incoming neighbours, perform some computation, potentially update the vertex state and then send messages through the outgoing edges of the vertex.
A vertex is the unit of parallelization and a vertex program receives a directed graph and a vertex function as input.
It was then extended to the concept of vertex scope, which includes the adjacent edges of the vertex.
The order of these steps will vary depending on the type of vertex-centric model used (scatter-gather, gather-apply-scatter).

\subsection{Model: Superstep Paradigm}\label{sec:computational-units:sec:superstep}

In a \textit{superstep} $S$, a user-supplied function is executed for each vertex $v$ (this can be done in parallel) that has a status of active.
When $S$ terminates, all vertices may send messages which can be processed by user-defined functions at step $S+1$.

\subsection{Model: Scatter-Gather}\label{sec:computational-units:sec:scatter-gather}

Scatter-gather shares the same idea behind vertex-centric but separates message sending from message collecting and update application~\cite{stutz2010signal}.
In the scatter phase, vertices execute a user-defined function that sends messages along outgoing edges.
In the gather phase, each vertex collects received messages and applies a user-defined function to update vertex state.

\subsection{Model: Gather-Apply-Scatter}\label{sec:computational-units:sec:gather-apply-scatter}

Gather-Sum-Apply-Scatter (GAS) was introduced by \texttt{PowerGraph}~\cite{Gonzalez:2012:PDG:2387880.2387883} and was aimed at solving the limitations encountered by vertex-centric or scatter-gather when operating on power-law graphs.
The discrepancy between the ratios of high-degree and low-degree vertices leads to imbalanced computational loads during a superstep, with high-degree vertices being more computationally-heavy and becoming stragglers.
GAS consists of decomposing the vertex program in several phases, such that computation is more evenly distributed across the cluster.
This is achieved by parallelizing the computation over the edges of the graph.
In the gather phase, a user-defined function is applied to each of the adjacent edges of each vertex in parallel.

\subsection{Unit: Edge-Centric (TLEV)}\label{sec:computational-units:sec:edge-centric}

The edge-centric approach, also referred as \textit{think-like-an-edge} (TLEV), was popularized by systems like \texttt{X\--Stre\-am}~\cite{Roy:2013:XEG:2517349.2522740} and \texttt{Cha\-os}~\cite{Roy:2015:CSG:2815400.2815408} which specify the computation from the point-of-view of edges.
These systems made of use of this paradigm to optimize the usage of secondary storage and network communication with cloud-based machines to process large graphs.

\subsection{Unit: Sub-graph-Centric (TLAG)}\label{sec:computational-units:sec:subraph-centric}

The previous models are subjected to higher communication overheads due to being fine-grained.
It is possible to use sub-graph structure to reduce these overheads (also known as \textit{component-centric}~\cite{heidari2018scalable}).
In this category, the work of \cite{kalavri2017high} denotes two sub-graph-centric approaches: \textit{par\-ti\-tion-centric} and \textit{nei\-gh\-bour\-hood-cen\-tric}. 
\textit{Partition-centric} instead of focusing on a collection of unassociated vertices, considers sub-graphs of the original graph.
Information from any vertex can be freely propagated within its physical partition, as opposed to the vertex-centric approach where a vertex only accesses the information of its most immediate neighbours.
This allows for reduction in communication overheads.
Ultimately, the partition becomes the unit of parallel execution, with each sub-graph being exposed to a user function.
This sub-graph-centric approach is also known as \textit{think-like-a-graph}~\cite{Tian:2013:TLV:2732232.2732238} (TLAG).
\textit{Neighborhood-centric}, on the other hand, allows for a physical partition to contain more than one sub-graph.
Shared state updates exchange information between sub-graphs of the same partition, with replicas and messages for sharing between sub-graphs that aren't in the same partition.
For completion, we refer the reader to an analysis of distributed algorithms on sub-graph centric graph platforms~\cite{kakwani2019distributed}.

\subsection{Model: MEGA}\label{sec:computational-units:sec:mega}

The MEGA model was introduced by \texttt{Tux2}~\cite{xiao2017tux2}, a system designed for graph computations in machine learning.
The model is composed of four functions defined by the user: an \textit{exchange} function which is applied to each edge and can change the value of the edge and adjacent vertices; an \textit{apply} function to synchronize the value of vertices with their replicas; a global \textit{sync} function to perform shared computations and update values shared among partitions; a \textit{mini-batch} function to indicate the execution sequence of other functions in each round.

There are graph processing systems that offer more than one type of model. 
To achieve parallelism and harness multiple machines in clusters, it is necessary to define how to break down the graph - we provide a high-level overview of methods employed in most well-known graph processing solutions

\section{Dimension: Partitioning}\label{sec:partitioning}

%

Graph partitioning is an important problem in graph processing, and this importance manifests in two formats.
The first, is out of a user's domain application with the goal of splitting the graph in parts which provide a relevant view of the data.
The second, is when partitioning may be considered as a \textit{hyper-algorithm}, that is, it is employed to divide the parts of the graph across a computational infrastructure, typically within the distributed systems' coordination layer, or across processing units or cores within machines.
Machine loads in distributed graph processing systems depend on the way computational units are distributed across workers.
The communication between them then depends on the number of units that are replicated.
We observe that partitioning has a cyclical nature to itself in the scope of distributed processing: one may wish to execute graph partitioning over a distributed system as part of a domain-specific problem; however, before that graph algorithm can execute, the graph data also incurs partitioning followed by distribution in the underlying (distributed) computational infrastructure.
While the study of graph partitioning is not recent, it gained additional depth in the last decade as the number of factors guiding optimization of partitioning increased with the complexity of graph processing systems.
We explore partitioning as a relevant dimension to classify graph processing systems as they must approach it in order to enable parallel computation over graphs.
The way it is approached becomes a distinctive feature between the systems.

Graph partitioning aims to divide the nodes of the graph into mutually-exclusive groups and to minimize the edges between partitions. 
This is effectively a grouping of the node set of the graph, which can represent a minimization of communication between partitions, with each partition for example assigned to a specific worker in a distributed system.
Partitioning is a task that produces groups of nodes, but grouping nodes is not only achieved with partitioning.
We note that other terms exist in the literature such as clustering and community detection. 
They are not interchangeable, for if a clustering algorithm breaks down the graph into three clusters, it does not necessarily hold true that each cluster represents its own community.
As an example, executing a clustering algorithm over a social network graph will result in a number of clusters. 
If each cluster represents for example a different continent, that does not necessarily mean each cluster represents one single community. 
Community detection algorithms, on the other hand, consider properties such as the density and interconnections within communities.
While clustering and community detection aim to identify similarities between nodes, their underlying assumptions of the graph are not equal, even though proposals have been made to map between these two tasks~\cite{guidotti2017equivalence}. 
Graph clustering shares similarities with graph partitioning in the sense that both produce groups of nodes. 
However, the objective functions they use are defined differently and subject to different constraints.
Graph partitioning, on which we focus, for example, requires that the number of groups (partitions) is known beforehand and is typically subject to more constraints.

\begin{figure*}
	\centering
	\subfigure[Edge-cut on $G$]
	{
		\resizebox{0.35\textwidth}{!}{
			\begin{tikzpicture}[show background rectangle,->,>=stealth',
	shorten >=1pt,auto,node distance=1.25cm,
    semithick,main node/.style={circle,draw}]

	\node[main node] (1) {1};
	\node[main node] (2) [left of=1] {2};
	\node[main node] (3) [below of=1] {3};
	\node[main node] (4) [below of=2] {4};
	\node[main node] (5) [right of=3] {5};
	\node[main node] (6) [right of=1] {6};

	\path
		(1) edge [bend right] node {} (2)
		(2) edge node [below] {} (1)
			  edge node [below] {} (4)
		(3) 
			edge node[right] {} (1)
			edge node[below] {} (2)
		(4) edge node [below] {} (3)
		edge node [below] {} (6)
		(6) edge node [below] {} (5)
		edge node [left] {} (1);
	\draw [->] [out=45,in=-45] (5) to (6);
	
	\draw [-] (-1.75,0.5) to (0.65,0.5);
	\draw [-] (-1.75,0.510506) to (-1.75,-1.7757);
	\draw [-] (-1.75,-1.73) to (0.65,-1.73);
	\draw [-] (0.62,0.510506) to (0.62,-1.7757);
	
	\draw [-] (0.80,0.5) to (1.78,0.5);
	\draw [-] (0.80,0.510506) to (0.80,-1.7757);
	\draw [-] (0.80,-1.73) to (1.78,-1.73);
	\draw [-] (1.75,0.510506) to (1.75,-1.7757);

\end{tikzpicture}
		}
		\label{fig:graph_partitioning:ec}
	}
	\subfigure[Edge-cut replication]
	{
		\resizebox{0.56\textwidth}{!}{
			\begin{tikzpicture}[show background rectangle,->,>=stealth',
	shorten >=1pt,auto,node distance=1.25cm,
    semithick,main node/.style={circle,draw}]

	\node[main node] (1) {1};
	\node[main node] (2) [left of=1] {2};
	\node[main node] (3) [below of=1] {3};
	\node[main node] (4) [below of=2] {4};
	\node[main node] (6) [right of=1] {6};
	\node[main node] (7) [right of=6] {1};
	\node[main node] (8) [below of=7] {4};
	\node[main node] (9) [right of=7] {6};
	\node[main node] (10) [below of=9] {5};

	\path
		(1) edge [bend right] node {} (2)
		(2) edge node [below] {} (1)
			  edge node [below] {} (4)
		(3) 
			edge node[right] {} (1)
			edge node[below] {} (2)
		(4) edge node [below] {} (3)
		edge node [below] {} (6)
		(6) edge node [left] {} (1)

		(8) edge node [below] {} (9)
		(9) edge node [below] {} (10)
		edge node [left] {} (7);

	\draw [->] [out=45,in=-45] (10) to (9);
	
	\draw [-] (-1.75,0.5) to (1.78,0.5);
	\draw [-] (-1.75,0.510506) to (-1.75,-1.7757);
	\draw [-] (-1.75,-1.73) to (1.78,-1.73);
	\draw [-] (1.75,0.510506) to (1.75,-1.7757);
	
	\draw [-] (2.0,0.5) to (4.28,0.5);
	\draw [-] (2.0,0.510506) to (2.0,-1.7757);
	\draw [-] (2.0,-1.73) to (4.28,-1.73);
	\draw [-] (4.25,0.510506) to (4.25,-1.7757);

\end{tikzpicture}
		}
		\label{fig:graph_partitioning:ec-repl}
	}
	\subfigure[Vertex-cut on $G$]
	{
		\resizebox{0.35\textwidth}{!}{
			\begin{tikzpicture}[show background rectangle,->,>=stealth',
	shorten >=1pt,auto,node distance=1.25cm,
    semithick,main node/.style={circle,draw}]

	\node[main node] (1) {1};
	\node[main node] (2) [left of=1] {2};
	\node[main node] (3) [below of=1] {3};
	\node[main node] (4) [below of=2] {4};
	\node[main node] (5) [right of=3] {5};
	\node[main node] (6) [right of=1] {6};

	\path
		(1) edge [bend right] node {} (2)
		(2) edge node [below] {} (1)
			  edge node [below] {} (4)
		(3) 
			edge node[right] {} (1)
			edge node[below] {} (2)
		(4) edge node [below] {} (3)
		edge node [below] {} (6)
		(6) edge node [below] {} (5)
		edge node [left] {} (1);
	\draw [->] [out=45,in=-45] (5) to (6);
	
	\draw [-] (-1.75,0.5) to (0.65,0.5);
	\draw [-] (-1.75,0.510506) to (-1.75,-1.7757);
	\draw [-] (-1.75,-1.73) to (-0.62,-1.73);
	\draw [-] (-0.6605,-0.55) to (0.65,-0.55);
	\draw [-] (0.62,-0.5394) to (0.62,-1.7757);
	
	\draw [-] (-0.65,0.5) to (1.78,0.5);
	\draw [-] (-0.65,0.510506) to (-0.65,-1.7757);
	\draw [-] (-0.65,-1.73) to (1.78,-1.73);
	\draw [-] (1.75,0.510506) to (1.75,-1.7757);

\end{tikzpicture}
		}
		\label{fig:graph_partitioning:vc}
	}
	\subfigure[Vertex-cut replication]
	{
		\resizebox{0.455\textwidth}{!}{
			\begin{tikzpicture}[show background rectangle,->,>=stealth',
	shorten >=1pt,auto,node distance=1.25cm,
    semithick,main node/.style={circle,draw}]

	\node[main node] (2) [left of=1] {2};
	\node[main node] (3) [below of=1] {3};
	\node[main node] (4) [below of=2] {4};

	\node[main node] (7) [right of=3] {3};
	\node[main node] (5) [right of=7] {5};
	\node[main node] (6) [above of=5] {6};
	\node[main node] (1) [above of=7] {1};

	\path
		(2) edge node [below] {} (4)
		(3) edge node[below] {} (2)
		(4) edge node [below] {} (3)
		(6) edge node [below] {} (5)
		edge node [left] {} (1)
		
		(7) edge node [above] {} (1);
		(6) edge node [below] {} (5);
	\draw [->] [out=45,in=-45] (5) to (6);
	
	\draw [-] (-1.75,0.5) to (0.55,0.5);
	\draw [-] (-1.75,0.510506) to (-1.75,-1.7757);
	\draw [-] (-1.75,-1.73) to (0.55,-1.73);
	\draw [-] (0.51,0.510506) to (0.51,-1.7757);
	
	\draw [-] (0.80,-1.73) to (3.03,-1.73);
	\draw [-] (0.80,0.521006) to (0.80,-1.7757);
	\draw [-] (0.80,0.510506) to (3.0457,0.510506);
	\draw [-] (3.00,0.510506) to (3.00,-1.7757);

\end{tikzpicture}
		}
		\label{fig:graph_partitioning:vc-repl}
	}
	\caption{Depiction of vertex-cut and edge-cut over the sample graph $G$.}
	\label{fig:partition-cut-models}
\end{figure*}
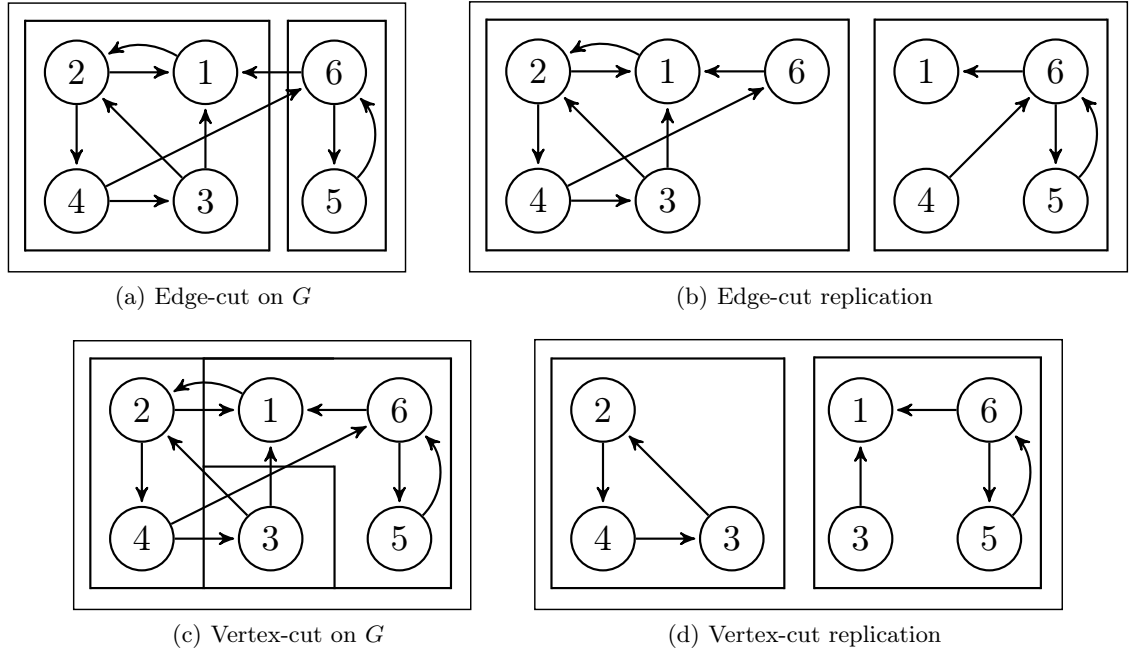

An earlier work on balanced graph partitioning~\cite{andreev2006balanced} defines the problem as $(k,v)$-balanced partitioning: to divide the vertices of the graph into $k$ components of almost equal size, with each of size less than $c\ \cdotp \frac{n}{k}$ for a given constant $c > 1$. 
It is a balanced $k$-way partitioning problem which has been studied in the literature~\cite{bulucc2016recent}.
To consider the partitioning problem as a challenge to enabling distributed processing, it is necessary to ask if the goal is to distribute the vertices (edge cut model - $EC$) or the edges (vertex cut model - $VC$) of the graph across machines in order perform it.
We provide detail into these problem formulations with an example of vertex-cut and edge-cut in Fig.~\ref{fig:partition-cut-models}.
Furthermore, different combinations between computational unit and cut model are possible: vertex-cut can be used to process in a vertex-centric~\cite{mofrad2018revolver} or edge-centric~\cite{Gonzalez:2012:PDG:2387880.2387883} way, and the same is possible using edge-cut used to partition a graph where computation is vertex-specific~\cite{bao2013towards,martella2017spinner} or edge-specific.

\subsection{Edge-Cut (EC)}\label{sec:partitioning:sec:ec}

Balanced $k$-way partitioning may be defined for edge-cut partitioning, which is associated to vertex-centric (TLAV) systems, the most common computational model in graph processing systems~\cite{Malewicz:2010:PSL:1807167.1807184,Gonzalez:2014:GGP:2685048.2685096,DBLP:books/sp/SOAK2016}.
We reproduce the definition of~\cite[Sec. 2]{soudani2019investigation} for this case, where for a given graph $G = (V, E)$, we wish to find a set of partitions $P = \{P_{1}, P_{2},\dots,P_{k}\}$. 
These partitions must be pairwise disjoint and their union is equal to $V$ while following these conditions~\cite{soudani2019investigation}:

\begin{equation}\label{eq:ec-partitioning-formulation}
\min_{P} | \{ e | e = (v_{i}, v_{j}) \in E, v_{i} \in P_{x}, v_{j} \in P_{y}, x \neq y\} | 
\end{equation}
\begin{equation}\label{eq:ec-partitioning-condition}
s.t. \dfrac{\max_{i}|P_{i}|}{\dfrac{1}{k} \sum_{i=1}^{k} |P_{i}|} \leq \epsilon
\end{equation}

Depending on the application objective for which this partitioning type will be performed, Eq.~\ref{eq:ec-partitioning-condition} should be adapted.
For example, in the case of machines having different characteristics, it should be considered that the load of any machine will be less than the maximum computing power.
Or if the graph structure is stored in secondary memory, the interest is on having balanced size partitions with high speed sequential storage access and decreasing the number of cut edges is no longer a focus.


\subsection{Vertex-Cut (VC)}\label{sec:partitioning:sec:vc}

In the vertex-cut model, the goal is to distribute edges across partitions.
They are placed in different partitions, with vertices being copied in partitions which have their adjacent edges.
Care must be taken to balance the number of edges per partition (its measure of size) and to minimize the number of vertex copies.
This objective may be formulated as such~\cite{soudani2019investigation}:

\begin{equation}\label{eq:vc-partitioning-formulation}
\min_{P} \dfrac{1}{|V|} \sum_{v \in V}{|P(v)|}
\end{equation}
\begin{equation}\label{eq:cv-partitioning-condition}
s.t. \max_{p_{i}} |\{ e \in E | P(e) = p_{i} \}| \leq \epsilon\dfrac{|E|}{k}
\end{equation}

Vertex-cut achieves better performance than edge-cut  for natural graphs such as those representing web structure and social networks~\cite{soudani2019investigation}.

\subsection{Hybrid-Cut (HC)}\label{sec:partitioning:sec:hc}

Hybrid strategies can be employed to perform the cuts.
They can for example be guided with heuristics such as vertex degree in order to decide what to do with them.
The \texttt{Pow\-er\-Ly\-ra}~\cite{chen2015powerlyra} system for example allocates the incoming edges of vertices with low degree in a worker.
It uses edge-cut for vertices of low-degree and vertex-cut for high-degree vertices.

\subsection{Stream-based Partitioning}\label{sec:partitioning:sec:stream-partitioning}

In these methods of partitioning, vertices or edges in the graph are analyzed in succession in a stream.
Placement decisions are made \textit{online}, that is, when the vertices or edges appear in the stream, and the decisions are based on the location of previous elements.
This is done under the assumption that there will be no information on the edges or vertices that will arrive in the flow of the stream.
This type of method can rely on edge-cut partitioning (e.g. \texttt{Random heuristic} and the \texttt{Linear Deterministic Greedy}~\cite{stanton2012streaming}, \texttt{Gemini} which uses chunk-based assuming adjacency list model~\cite{zhu2016gemini}, \texttt{Fennel}~\cite{tsourakakis2014fennel}), vertex-cut partitioning (e.g. \texttt{Grid} heuristic~\cite{jain2013graphbuilder}, \texttt{PowerGraph} greedy heuristic~\cite{Gonzalez:2012:PDG:2387880.2387883}, \texttt{Graphbuilder}~\cite{jain2013graphbuilder} placing the edge in the smallest partition, \texttt{HDRF}~\cite{petroni2015hdrf} method which takes into consideration vertex degrees) and there are aspects of these methods that will have different approaches regarding how this is achieved with parallel and distributed execution.
Stream-based partitioning is also used as a good choice for loading the graph as it does not have to be fully loaded in memory for partitioning.

\subsection{Distributed Partitioning}\label{sec:partitioning:sec:distributed-partitioning}

Many distributed partitioning algorithms are based on label propagation~\cite{Boldi:2011:LLP:1963405.1963488,gregory2010finding,liu2010advanced,zhur2002learning}, with variations such as how the specific labeling of a vertex should be influenced by its neighbours, if it should also be influenced by the label's global representation in the graph and also constraints on the minimum and maximum sizes required for partitions.
For example, \texttt{Revolver}, which performs vertex-centric graph partitioning with reinforcement learning, assigns an agent to each vertex, with agents assigning vertices to partitions based on their probability distribution (these are then refined based on feedbacks~\cite{mofrad2018revolver}).
The authors of~\cite{soudani2019investigation} note that other approaches consider the partitioning problem as a multi-objective and multi-constraint problem, achieving better results compared to one-phase methods~\cite{slota2014pulp}.
Distributed partitioning systems are good for when partitioning is performed once and then calculations are repeatedly performed.

\subsection{Dynamic Graph Partitioning}\label{sec:partitioning:sec:dynamic-partitioning}

When the graph is no longer static, vertex and edges may be added or removed as time passes - this is especially true in social networks.
This implies that for graphs from which we need to perform computations as they evolve, the original partitioning may become inefficient.
With predictable algorithm runtime characteristics, it becomes feasible to keep close the vertices which will be used together in the same supersteps, using for example graph traversal algorithms.
But when this is not the case, systems can be designed for example to monitor the load and communication of the machines and migrate vertices as appropriate, with different techniques having been proposed for that purpose (among others, \texttt{xDGP}~\cite{vaquero2013xdgp} to repartition massive graphs to adapt to structural changes, \texttt{GPS}~\cite{salihoglu2013gps} which reassigns vertices based on communication patterns, \texttt{X-Pregel}~\cite{bao2013towards} with reduction of message exchanges and dynamic repartitioning).
Dynamic partitioning methods have the advantage of outputting very good load balancing and communication cost reductions due to considering heterogeneous hardware and runtime characteristics.

\subsection{Partitioning: Summary}\label{sec:partitioning:sec:state-of-art}

Employed graph partitioning strategies vary, with different systems offering different solutions.
Among performance-impacting factors \cite{soudani2019investigation}, we have the number of active vertices and edges influencing machine load.
At the same time, communication will be more expensive depending on how replication of edges and vertices is performed.
Partitioning must balance communication and machine loads.
The partitioning challenge in vertex-centric systems is relevant due to how widespread this model is.
The authors of \cite{soudani2019investigation} note three major approaches for big graph partitioning: \textit{a)} partitioning the graph serially in a single pass and permanently assigning the partition on the first time an edge or vertex is assigned (stream-based); \textit{b)}; methods that partition in a distributed way; \textit{c)} dynamic methods that adapt the partitions based on monitoring the load and communication of machines during algorithm execution.
The way the distribution is achieved and data is represented will be a factor in going beyond the \textit{read-eval-write loop}.
In this scope, a dynamic method would be necessary as a basis to develop the properties we described.
For an in-depth analysis of partitioning methods, vertex cut models and their relation to the dynamic nature of data, we invite the reader to read~\cite{soudani2019investigation}.

Being able to decompose the graph is a cornerstone for efficient and distributed computation of graphs. 
An equally-important aspect that determines how we must approach the computation is the possible dynamism of the graph. 
A static graph over which we want to perform analytics is a scenario different from maintaining a large graph available for separate queries and susceptible to updates.

\section{Dimension: Dynamism}\label{sec:dynamism-evolving-temporal-graphs}

We include and consider \textit{dynamism} a relevant dimension of graph processing due to there existing different meanings associated to it in the literature and for which different systems can be attributed.
While one may consider static graphs to be completely unrelated to dynamism, there is in fact a relation to it due to what is known as \textit{stream processing}.
For example, a graph processing system may ingest an unbounded stream of edges and update statistics over the stream (e.g., keeping a triangle count updated~\cite{Ahmed:2017:SMG:3137628.3137651}), but stream processing may also take place in static graph processing.
This is the case with approaches that process a static graph but process its elements from a stream perspective (e.g., \texttt{Cha\-os}~\cite{Roy:2015:CSG:2815400.2815408} and \texttt{X\--Stre\-am}~\cite{Roy:2013:XEG:2517349.2522740} with their edge-centric approach).
Considering if a system targets graphs that change or are immutable (static) is an obvious way to separate graph processing systems when classifying them.
However, this dimension is actually a spectrum between the immutable (e.g. stream-based perspectives to process static graphs) and the changing - for example, is the whole graph structure kept in memory (or secondary storage) in a single machine (or across cluster nodes), or is it discarded by proxy of some criteria (and thus one simply updates mathematical properties of the graph using only recent information from the stream)?
For this spectrum, the authors of~\cite{besta2019practice} cover definitions found in the literature:

\begin{itemize}

\item Temporal graphs. 
These are, in essence, static graphs which have annotated temporal information which allows for recreating the domain represented by the graph at any given point in time.
It is not structurally-changed while doing so; it means that for a given time range or event, only the elements with valid timestamps under required constraints are considered for computation.
The work of~\cite{kostakos2009temporal} introduces the temporal graph as a representation encoding temporal data into the graph while retaining the temporal information of the original data.
They present metrics that can be used to study temporal graphs and use the representation to explore dynamic temporal properties of data using graph algorithms without requiring data-driven simulations.
\texttt{ImmortalGraph}~\cite{miao2015immortalgraph} is a storage and execution engine designed with temporal graphs in mind, having achieved grea\-ter efficiency than database solutions for graph queries. 
\texttt{ImmortalGraph} schedules common bu\-lk operations in a way to maximize the benefit of in-memory data locality.
It explores the relation between locality, parallelism and incremental computation while enabling mining tasks on temporal graphs.
For more information and reach on the topic of temporal graphs, we direct the reader to~\cite{michail2016introduction}.

\item Streaming graph algorithms~\cite{feigenbaum2005graph}.
With these, the common scenario starts from an empty graph without edges (and a fixed set of vertices).
For each algorithm step, a new edge is inserted into the graph or an edge is removed.
It is desired that these algorithms are developed to minimize parameters such as graph data structure storage, the time to process an edge or the time to recover the final solution. 
There exist several systems which process streaming graph computations - we note also for the reader a recent framework for comparing the systems aimed at this type of dynamism~\cite{erb2018graphtides}.
The \texttt{STINGER} data structure has been used for streaming graphs as well ~\cite{ediger2012stinger}.

\item Sketching and dynamic graph streams.
Sketching techniques~\cite{Ahn:2012:GSS:2213556.2213560} may be applied to the edge incidence matrix of the input graph to approximate cut structure and connectivity.
The idea is to consume a stream of events in order to generate a probabilistic data structure representing properties of the graph.

\item Multi-pass streaming graph algorithms.
In this type of algorithm, all updates are streamed more than once in order to approximate the computation quality of the solution.
Additional complexity can emerge on how the strea\-ming model behaves - it can for example allow for the stream to be manipulated across passes~\cite{aggarwal2004streaming} or to stream sorting passes~\cite{demetrescu2009trading}.

\item Dynamic graph algorithms.
For these types, the focus is cast on being able to approximate combinatorial properties of the graph~\cite{besta2019practice} (e.g., connectivity, shortest path distance, cuts, spectral properties) while processing insertions and deletions.
The objective with this type of algorithm is to quickly integrate graph updates.
\texttt{Rin\-go}~\cite{Perez:2015:RIG:2723372.2735369} is a single-machine analytics system that supports dynamic graphs.

\end{itemize}

While partitioning and dynamism are relevant aspects, the scope of graph processing solutions in both industry and academia was shaped by the type of executed workloads.

\section{Dimension: Workload}\label{sec:workload-type}

The type of workload performed by a graph processing system also plays an important role in classifying them. 
The type of task performed by graph databases is different from the systems that run global algorithms over them.
The concept of analyzing a graph takes on different contexts depending on user needs.
We note that when a graph is to be \textit{processed}, the scope of its data analysis usually falls in these two categories:

\begin{enumerate}[label=(\alph*)] 

\item To retrieve instances of domain-specific relations in the graph (e.g. pattern matching, multi-hop queries). 
These are usually found in graph databases, with an emphasis on optimization of data query and storage for online transaction processing scenarios.
This is often accompanied with the use of graph query languages (GQLs) to execute queries that return a view on the graph and also potentially producing effects on it. 

\item To execute an algorithm over the whole graph (e.g. PageRank, connected components, detecting communities, finding shortest paths). 
The solutions for this task, performance-wise, aim to achieve high-performance computational throughput, whether using distributed systems or a single-machine setup. 
It is a focus leaning on the data analytics aspect. 
\end{enumerate}

The former (a) is a common scenario in graph databases such as \texttt{Neo4j}~\cite{Webber:2012:PIN:2384716.2384777} and \texttt{Ja\-nus\-Gra\-ph}~\cite{janus}, among others.
These databases offer graph query languages (usually even allowing interchangeability between languages) such as \texttt{Cy\-pher} or \texttt{Grem\-lin}~\cite{Holzschuher:2013:PGQ:2457317.2457351}.
They are built to store the graph, some with sharding (horizontal scaling) to distribute the graph across the storage/computational infrastructure (some outsource the storage medium to da\-ta\-ba\-se technologies such as \texttt{HBase}~\cite{DBLP:books/daglib/0027893} or \texttt{Ca\-s\-san\-dra}~\cite{Lakshman:2010:CDS:1773912.1773922}), others in a centralized server (but allowing cluster nodes for the specific purpose of redundancy).
They employ schemes to store the graph efficiently while offering transaction mechanisms to operate over the graph and to perform queries. 

The latter type (b) is seen in big (graph) data processing systems like \texttt{Spark (GraphX library)}~\cite{graphx} and \texttt{Flink (Gelly library)}~\cite{carbone2015apache}.
The mentioned names are all distributed processing frameworks that can take advantage of multi-core machines and clusters. 
These systems and their libraries allow for expressive computation over graphs in few lines of code.
Many of the systems come with their sets of graph algorithms, allowing for the composition of workflows while abstracting away many details from the programmer (regarding distributed computation orchestration and the internal implementation of the graph algorithms).

It is important to consider two definitions regarding the nature of computational tasks: online analytical processing (\texttt{OLAP}) and online transaction processing (\texttt{OLTP}).
\texttt{OLAP} is an approach to enable answering multi-dimensional analytical queries quickly.
Among its instances we may find tasks such as business reporting for sales, management reporting, business process management~\cite{benisis2010business}, financial reporting and others. 
\texttt{OLTP}, on the other hand, refers to systems that enable and manage transaction-oriented applications, with \textit{transaction} meaning in a computational context the atomic state changes that take place in database systems.
\texttt{OLTP} examples include retails sales and financial transaction systems, and applications of this type tend to be high-throughput and update/insertion-intensive in order to provide availability, speed, recoverability and concu\-rren\-cy\-\cite{oltp_oracle}.

The earlier type $a)$ of graph processing task may be associated to \texttt{OLTP} systems as the goal is to store representations of graphs by quickly ingesting new information, efficiently representing it regarding space consumption and access speed, and being able to execute updates under ACID properties (or a subset of those).
For this type of task $a)$, one may find numerous graph databases to match the description, such as those for designed for semantic (\texttt{RDF}) representations~\cite{ontotext_graphdb,stutz2013triplerush,tesoriero2013getting,allegrograph}, or for property graph models~\cite{graphengine,miller2013graph,kankanamge2017graphflow,cailliau2019redisgraph,rudolf2013graph,titan,janus,datastax,paz2018introduction,martinez2011dex,hwang2018graph,DBLP:journals/pvldb/DubeyHES16,arangodb_github}, both~\cite{bebee2018amazon} and also other specific purposes~\cite{cayley_github}.
The latter type of task $b)$ may be associated to \texttt{OLAP}, where there is a focus on extracting value from the data and the nature of the task is typically read-only.
We include graph processing systems (not databases) in this group of \texttt{OLAP}-type tasks, even the systems which support mutability in graphs due to supporting dynamism in any form.

There is a considerable overlap between \texttt{OLTP}-type tasks and graph databases, and there is also an overlap between \texttt{OLAP}-type tasks and graph processing systems.
While the distinction between \texttt{OLAP} and \texttt{OLTP} task types is not a dimension that perfectly divides systems in the graph processing landscape, we note that such a distinction holds value in guiding future taxonomies of the graph processing system landscape, and for that reason we include it as a dimension.

The way these three dimensions are accounted for influence the design of graph processing systems. 
Many different architectures exist, for which we share an exhaustive list of specific solutions, from single-machine systems to parallel processing in clusters and storage in tailor-made graph databases.

\section[Single-Machine and Shared-Memory Parallel Approaches]{Single-Machine and Shared-Memory Parallel Approa-\\ches}\label{sec:single-machine}

\begin{itemize}

\item \texttt{Gra\-ph\-Lab}~\cite{Low_graphlab:a} was published as a framework (implemented in \texttt{C++}) for parallel machine learning and later extended to support distributed settings while retaining strong data consistency guarantees~\cite{DBLP:journals/corr/LowGKBGH14}. 
The authors evaluate it on Amazon EC2, outperforming equivalent \texttt{Map\-Re\-du\-ce} implementations by over 20X and ma\-tch the performance of specifically-crafted \texttt{MPI} implementations. 
\texttt{Gra\-ph\-Lab} requires the whole graph and program state to reside in RAM.
It uses a chromatic engine so that no adjacent vertices have the same color and to enable the efficient use of network bandwith and processor time. 
The authors evaluate it for applications such as Netflix movie recommendation, video co-segmentation and named entity recognition. 
It is open-source~\cite{graphlab_github} under the \texttt{A\-pa\-che} \texttt{Li\-cen\-se} \texttt{2.0}.

\item \texttt{GRA\-CE}~\cite{wang2013asynchronous} is a synchronous iterative graph programming model, with separation of application logic and execution policies.  
Its design includes the implementation (\texttt{C++}) of a parallel execution engine for both synchronous and user-specified asynchronous execution policies.
\texttt{GRA\-CE} stores directed graphs, and in its model and the computation is expressed and per\-for\-med in a way similar to \texttt{Pre\-gel}. 
It provides additional flexibility, by allowing the user to relax synchronization of computation. 
This is achieved with user-defined functions which allow updating the scheduling priority of vertices that receive messages (the vertex order in which computation will take place within an iteration). 
\texttt{GRA\-CE}'s design targets both shared-memory and distributed system scenarios, but the initial prototype focuses on shared-memory.
We did not find the source code available.

\item \texttt{Li\-gra}~\cite{shun2013ligra} is a \texttt{C++} lightweight graph processing framework targeting shared-memory para\-l\-lel\-/mul\-ti\--co\-re machines, easing the writing of graph traversal algorithms. 
This framework offers two map primitives to operate a given logic on vertices (\texttt{Ver\-tex\-Map}) and edges (\texttt{Edge\-Map}) and supports two data types: the traditional graph $G = (V, E)$ as we described in an earlier section, and another one to represent subsets of vertices. 
The interface is designed to enable the processing of edges in different orders depending on the situation (as opposed to \texttt{Pre\-gel} or \texttt{Gi\-ra\-ph}). 
The code of \texttt{Li\-gra} represents in-edges and out-edges as arrays, with in-edges for all vertices being partitioned by their target vertex and storing the source vertices, and the out-edges are in an array partitioned by source vertices and storing the target vertices.
While the authors claim to have achieved good performance results, they mention \texttt{Li\-gra} does not support algorithms based on modifying the input graph. 
It is available~\cite{ligra_github} under the \texttt{MIT License}.

\item \texttt{Rin\-go}~\cite{Perez:2015:RIG:2723372.2735369} is an approach for multi-core single-machine big-memory setups.
It is a high\--per\-for\-mance interactive analytics system using a \texttt{Py\-thon} front-end on a scalable parallel \texttt{C++}  back-end, representing the graph as a hash table of nodes. 
It supports fast execution times with exploratory and interactive use, offering graph algorithms in a high-level language and rich support for transformations of input data into graphs.
\texttt{Rin\-go} is open-source and available~\cite{ringo_github} under the \texttt{BSD License}.

\item \texttt{Po\-ly\-mer}~\cite{zhang2015numa} is a NUMA-aware graph analytics system on multi-core machines that is open-source~\cite{polymer_github} under the \texttt{A\-pa\-che} \texttt{Li\-cen\-se} \texttt{2.0} and implemented in \texttt{C++}. 
It innovated by differentially allocating and placing topology data, application-defined data and mutable run-time graph system states according to access patterns to minimize remote accesses. 
\texttt{Po\-ly\-mer} also deals with random accesses by converting the random ones into sequential remote accesses using lightweight vertex replication across the NUMA nodes. 
It was built with a hierarchical barrier for increased parallelism and locality. 
The design also includes edge-oriented balanced partitioning for skewed graphs and adaptive data structures in function of the fraction of active vertices. 
It was compared to \texttt{Li\-gra}, \texttt{X\--Stre\-am} and \texttt{Galois} on an 80-core Intel machine (no hyper-threading) and on a 64-core AMD machine. 
For different algorithms across several data sets, \texttt{Po\-ly\-mer} consistently almost always achieved the lowest execution time.

\item \texttt{Gra\-ph\-Mat}~\cite{sundaram2015graphmat} is a framework written in \texttt{C++} aimed at bridging the user-friendly graph analytics and native hand-optimized code. 
It presents itself as a vertex-centric framework without sacrificing performance, as it takes vertex programs and maps them to exclusively use sparse matrix high-performance back-end operations. 
\texttt{Gra\-ph\-Mat} takes graph algorithms expressed as vertex programs and performs generalized sparse matrix vector multiplication on them.
It achieved greater performance than other frameworks such as 5-7X faster than \texttt{Gra\-ph\-Lab}, \texttt{Galois} and  \texttt{ComBLAS}. 
It also achieved multi-core scalability, being over 10X faster than single-threaded implementation on a 24-core machine. 
It is open-source and available\-\cite{graphmat_github} under specific conditions by Intel.

\item \texttt{Mo\-saic}~\cite{Maass:2017:MPT:3064176.3064191} is a system for single heterogeneous machines with fast storage media (e.g., NVMe and SSDs) and massively-parallel co-processors (e.g., Xeon Phi) developed to enable the processing of trillion-edge graphs.
The system is designed explicitly separating graph processing engine components into scale-up and scale-out goals.
It is written in \texttt{C++} uses a compact representation of the graph using Hilbert-ordered tiles for locality, load balancing and compression and uses a hybrid computation model that uses both vertex-centric operations (on host processors) and edge-centric operations (on co-processors). 
\texttt{Mo\-saic} is open-sour\-ce\-\cite{mosaic_github} under the \texttt{MIT License}.

\end{itemize}

\section[High-Performance Computing]{High-Performance Compu-\\ting}\label{sec:relevant_hpc_systems}

These systems are hallmarks of high-performance computing solutions applied to graph processing.
Their merits encompass algebraic decomposition of the major graph operations, implementing them and translating them across different homogeneous layers of parallelism (across cores, across CPUs).
Here we mention what are, to the best of our knowledge, the most relevant works:

\begin{itemize}

\item \texttt{Parallel Boost Graph Library} (\texttt{P\-B\-G\-L})~\cite{gregor2005parallel} .
It is an extension (\texttt{C++}) of \texttt{Boost's} graph library.
It is a distributed graph computation library, also offering abstractions over the communication medium (e.g. \texttt{MPI}).
The graph is represented as an adjacency list that is distributed across multiple processors.
In \texttt{P\-B\-G\-L}, vertices are divided among the  processors, and each vertex's outgoing edges are stored on the processor storing that vertex. 
\texttt{P\-B\-G\-L} was evaluated on a system composed of 128 compute nodes connected via Infiniband.
It is available~\cite{PBGL_github} under a custom \texttt{Boost} \texttt{Software} \texttt{Li\-cen\-se} \texttt{1.0}.

\item \texttt{Com\-b\-BLAS}~\cite{bulucc2011combinatorial}.
A parallel graph distributed-memory library in \texttt{C++} offering  linear algebra primitives based on sparse arrays for graph analytics.
This system considers the adjacency matrix of the graph as a sparse matrix data structure.
\texttt{Com\-b\-BLAS} is edge-based in the sense that each element of the matrix represents an edge and the computation is defined over it.
It decouples the parallel logic from the sequential parts of the computation and makes use of \texttt{MPI}.
However, its \texttt{MPI} implementation does not take advantage of flexible shared-memory operations.
Its authors targeted hierarchical parallelism of supercomputers for future work.
It is available~\cite{comblas_repo} under a custom license.

\item \texttt{Ha\-vo\-q\-GT}~\cite{pearce2013scaling} is a \texttt{C++} system with techniques for processing scale-free graphs using dis\-tri\-bu\-ted memory.
To handle the scale-free properties of the graph, it uses edge list partitioning to deal with high-degree vertices (hubs) and dummy vertices to represent them to reduce communication hot spots.
\texttt{Ha\-vo\-q\-GT} allows algorithm designers to define vertex-centric procedures in a distributed asynchronous visitor queue.
This queue is part of an asynchronous visitor pattern designed to tackle load imbalance and memory latencies.
\texttt{Ha\-vo\-q\-GT} targets supercomputers and clusters with local NV\-RAM. 
It is available~\cite{havocgt_repo} under the \texttt{GNU} \texttt{Les\-ser} \texttt{Ge\-ne\-ral} \texttt{Public} \texttt{License 2.1}.

\end{itemize}

\section[Distributed Graph Processing Systems]{Distributed Graph Proces-\\sing Systems}\label{sec:distributed-systems}

While the previous systems we detailed performed analytics and enabled the execution of graph algorithms, they did so with a focus on specific hardware and distributed memory.
We list here some of the most relevant state-of-the-art systems used for graph processing in the scope of analytics (\texttt{OLAP}). 
Their use of different architectures (from using local commodity clusters to cloud-based execution) and greater flexibility of deployment scenarios differentiate them from those of the previous section.
The following systems are relevant names in the literature, with \texttt{Gi\-ra\-ph} being the first open-source implementation of the \texttt{Pre\-gel} approach to graph processing, and \texttt{Spark} and \texttt{Flink} being open-source general distributed processing systems with graph processing APIs:

\begin{itemize}

\item \texttt{A\-pa\-che} \texttt{Gi\-ra\-ph}~\cite{ching2013scaling} is an open-source \texttt{Ja\-va} implementation of \texttt{Pre\-gel}~\cite{Malewicz:2010:PSL:1807167.1807184}, tailor-made for graph algorithms, supporting the GAS model and licensed~\cite{giraph_github} under the \texttt{A\-pa\-che} \texttt{Li\-cen\-se} \texttt{2.0}.
It was created as an efficient and scalable fault-tolerant implementation on clusters with thousands of commodity hardware, hiding implementation details underneath abstractions.
Work has been done to extend \texttt{Gi\-ra\-ph} from the \textit{think-like-a-vertex} (TLAV) model to \textit{think-like-a-graph} (TLAG)~\cite{Tian:2013:TLV:2732232.2732238}.
It uses \texttt{Hadoop}'s \texttt{MapRedu\-ce} implementation to process graphs.
\texttt{Gi\-raph}~\cite{giraph_github} allows for master computation, sh\-ar\-ded aggregators (relevant when computing a final result comprised of intermediate data from nodes), has edge-oriented input, and also uses out-of-core computation -- limited partitions in memory.
Partitions are stored in local disks, and for cluster computing settings, the out-of-core partitions are spread out across all disks.
\texttt{Giraph} attempts to keep vertices and edges in memory and uses only the network for the transfer of messages.
Improving \texttt{Gi\-ra\-ph}'s performance by optimizing its messaging overhead has also been studied~\cite{Liu:2016:GSO:2983323.2983726}.
It is interesting to note that single-machine large-memory systems such as \texttt{Rin\-go} highlight the message overhead as one of the major reasons to avoid a distributed processing scheme.

\item \texttt{Nai\-ad} is an open-source~\cite{naiad_github} (\texttt{A\-pa\-che} \texttt{Li\-cen\-se} \texttt{2.0}) dataflow processing system~\cite{Murray:2013:NTD:2517349.2522738} offering different levels of complexity and abstractions to programmers.
It allows programmers to implement graph algorithms such as weakly connected components,  approximate shortest paths and others while achieving better performance than other systems.
\texttt{Nai\-ad} is implemented in \texttt{C\#} and allows programmers to use common high-level APIs to express algorithm logic and also offers a low-level API for performance.
Its concepts are important and other systems could benefit from offering tiered programming abstraction levels as in \texttt{Nai\-ad}. 
Its low-level primitives allow for the combination of dataflow primitives (similar to those \texttt{VeilGraph} uses from \texttt{Flink}) with finer-grained control on iterative computations.
An extension to \texttt{Flink}'s architecture to offer this detailed control would enrich the abilities that our framework is able to offer to users. 

\item \texttt{A\-pa\-che} \texttt{Flink}~\cite{carbone2015apache}, formerly known as \texttt{Stratosphere} \cite{Alexandrov2014}, it is a framework which supports built-in iterations~\cite{carbone2015apache} (and delta iterations) to efficiently aid in graph processing and machine learning algorithms.
It is licensed~\cite{flink_github} under the \texttt{A\-pa\-che} \texttt{Li\-cen\-se} \texttt{2.0} and has a graph processing API called \texttt{Gelly}, which comes packaged with algorithms such as PageRank, single-source shortest paths and community detection, among others. 
\texttt{Flink} supports \texttt{Ja\-va}, \texttt{Py\-thon} and \texttt{Scala}.
It explicitly supports three vertex-based programming models: \textit{think-like-a-vertex} (TLAV) described as the most generic model, supporting arbitrary computation and messaging for each vertex; Sca\-tter\--Ga\-ther, whi\-ch separates the logic of message production from the logic of updating vertex values, which may typically make these programs have lower memory requirements (concurrent access to the inbox and outbox of a vertex is not required) while at the same time potentially leading to non-intuitive computation patterns; Gather-Sum-Apply-Scatter (GAS), which is similar to Scatter-Gather but the Gather phase parallelizes the computation over the edges, the messaging phase distributes the computation over the vertices and vertices work exclusively on neighbourhood, where in the previous two models a vertex can send a message to any vertex provided it knows its identification.
It supports all \texttt{Hadoop} file systems as well as \texttt{Amazon S3} and \texttt{Google Cloud} storage, among others.
Delta iterations are also possible with \texttt{Flink}, which is quite relevant as they take advantage of computational dependencies to improve performance.
It also has flexible windowing mechanisms to operate on incoming data (the windowing mechanism can also be based on user-specific logic).
Researchers have also looked into extending its \texttt{DataStream} constructs and its streaming engine to deal with applications where the incoming flow of data is gra\-ph-ba\-sed\cite{gelly-streaming}.

\item \texttt{A\-pa\-che} \texttt{Spark}~\cite{Zaharia:2010:SCC:1863103.1863113} and its \texttt{GraphX}~\cite{graphx} graph processing library, licensed~\cite{spark_github} under the \texttt{A\-pa\-che} \texttt{Li\-cen\-se} \texttt{2.0}.
It is a graph processing framework built on top of \texttt{Spark} (a framework supporting \texttt{Ja\-va}, \texttt{Py\-thon} and \texttt{Scala}), enabling low-cost fault-tolerance.
The authors target graph processing by expressing graph-specific optimizations as distributed join optimizations and graph views' maintenance.
In \texttt{GraphX}, the property graph is reduced to a pair of collections.
This way, the authors are able to compose graphs with other collections in a distributed dataflow framework.
Operations such as adding additional vertex properties are then naturally expressed as joins against the collection of vertex properties.
Graph computations and comparisons are thus an exercise in analyzing and joining collections.

\item \texttt{Gra\-ph\-Tau}~\cite{iyer2016time} is a time-evolving graph processing framework on top of \texttt{Spark} (\texttt{Ja\-va}, \texttt{Sca\-la}). 
It represents computations on time evolving graphs as a stream of consistent and resilient graph snapshots and a small set of operators that manipulate such streams. 
\texttt{Gra\-ph\-Tau} builds fault-tolerant graph snapshots as each small batch of new data arrives. 
It is also able to periodically load data from graph databases and reuses many operators from \texttt{GraphX} and \texttt{Spark Streaming}. 
For algorithms (based on label propagation) that are not resilient to gra\-ph changes, \texttt{Gra\-ph\-Tau} introduced an online rectification model, where errors caused by underlying graph modifications are corrected in online fashion with minimal state.
Its API frees programmers from having to implement graph snapshot generation, windowing operators and differential computation mechanisms.
We did not find its source code available.

\item \texttt{Tink}~\cite{lightenberg2018tink} is a library for distributed temporal graph analytics. 
It is built on \texttt{Flink} (\texttt{Ja\-va}, \texttt{Scala}) and focuses on \textit{interval} graphs, where each edge has an associated starting time and ending time. 
The author created different gra\-phs with information provided by Facebook and Wikipedia in order to evaluate the framework. 
\texttt{Tink} defines a temporal property graph model.
It is available online~\cite{tink_github}, although we did not find information pertaining licensing.

\end{itemize}

To the best of our knowledge, both \texttt{Flink} and \texttt{Spark} are currently the most widely-known distribu\-ted processing frameworks (we note \texttt{Gra\-ph\-Tau}, although its code is not available, is built over \texttt{Spark}) based on dataflow programming.
While the use of data\-flows grants flexibility to program implementation and execution by decoupling the program logic from how it is translated to the workers of a cluster, the graph libraries of these systems do not allow in an efficient way for a graph to be updated using stream-processing semantics while also maintaining the graph structure during computation.
It is possible to update graphs using these systems, but they make use of batch processing APIs for which the dataflow graphs must not become excessively big (or else dataflow plan optimizers may be \textit{locked} in the phase of exploring the optimization space of the execution plan) and graph must be periodically written to secondary storage (as a solution to avoid having progressively bigger execution plans).

\texttt{Flink}'s \texttt{Gelly} library has been used in \texttt{GRA\-DO\-OP}, which is an open-source~\cite{gradoop_github} (\texttt{A\-pa\-che} \texttt{Li\-cen\-se} \texttt{2.0}) distributed graph analytics research framework~\cite{DBLP:journals/pvldb/JunghannsKTGPR18} under active development and providing higher-level operations. 
\texttt{GRA\-DO\-OP} extends \texttt{Gelly} with additional specialized operators such as a graph pattern matching operator (which abstracts a cost-based query engine) and a graph grouping operator (implemented as a composition of map, filter, group and join transformations on \texttt{Flink}'s \texttt{DataSet}).
\texttt{GRA\-DO\-OP} also adopts the \texttt{Cy\-pher} query language (typically found in graph databases like \texttt{Neo4j}) to express logic that is translated to the relational algebra that underlies \texttt{Flink}'s \texttt{DataSet}~\cite{DBLP:conf/grades/JunghannsKAPR17}. 

In a similar way, \texttt{Spark} has its graph processing library \texttt{Graphx} which was built over the system's batch processing API, like the case of \texttt{Flink}'s \texttt{Gelly} and also suffering from the same previously mentioned limitations.
A higher-level API was designed to extend the functionalities of \texttt{GraphX} while harnessing \texttt{Spark}'s \texttt{DataFrame} API.
For this, the \texttt{GraphFrames} open-source~\cite{graphframes_github} (\texttt{A\-pa\-che} \texttt{Li\-cen\-se} \texttt{2.0}) library was created~\cite{dave2016graphframes}. 
A look at its implementation reveals that it has less high-level operations than \texttt{Gelly}.
Effectively, without simulating some of \texttt{Gelly}'s API, equivalent programs in \texttt{GraphX} lend themselves to more conceptual verbosity due to the lack of syntactic sugar. 

\begin{figure}[h]
	\centering
	\includegraphics[width=0.5\textwidth]{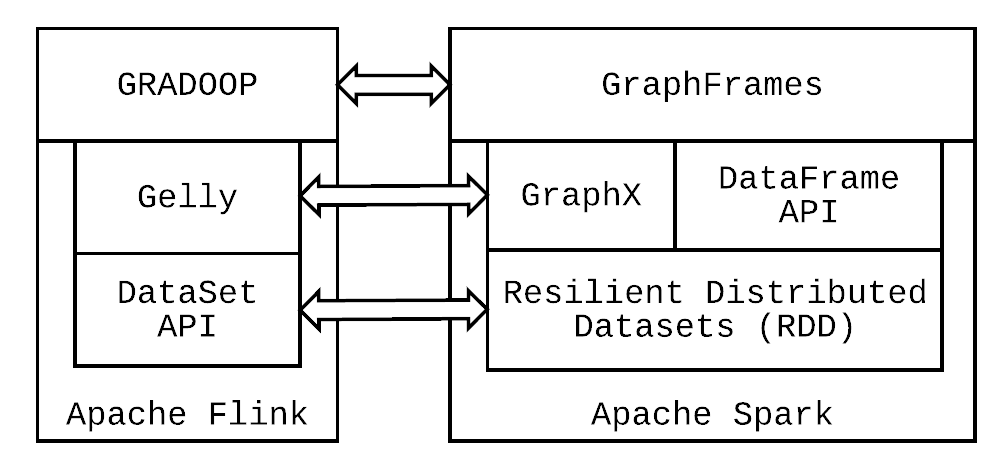}
	\caption{Contrast of the \texttt{Flink} and \texttt{Spark} distributed dataflow ecosystems for graph processing.}
	\label{fig:gradoop_vs_graphframes}
\end{figure}

We display in Figure~\ref{fig:gradoop_vs_graphframes} parallels between \texttt{Flink, Spark} and the graph processing ecosystems built on top of them.  
\texttt{Gelly}'s equivalent in \texttt{Spark} is \texttt{GraphX}, implemen\-ted in \texttt{Scala}.
Vertices and edges are manipulated by using \texttt{Spark}'s \texttt{Resilient Distributed Datasets} (\texttt{RDDs}), which can be viewed as a conceptual precursor to \texttt{Flink}'s \texttt{DataSet}.
\texttt{Spark} also offers the \texttt{DataFrame} API to enable tabular manipulation of data.
\texttt{Graph\-Frames} is another graph processing library for \texttt{Spark}.
While it has interoperability and a certain overlap with the functionality offered in \texttt{GraphX}, it integrates the tabular perspective supported by \texttt{Spark}'s \texttt{DataFrame} API and also supports performing traver\-sal-like queries of the graph via \texttt{SparkSQL}.
In this way, \texttt{GraphFrames} provides graph analytics capabilities in \texttt{Spark} much the same way \texttt{GRA\-DO\-OP} does in \texttt{Flink}.

The next two examples, \texttt{X\--Stre\-am} and \texttt{Cha\-os} are grouped together as they brought relevance to the edge-centric (TLAE) model and employed it to explore novel ways to balance network latencies and use of SSDs to increase performance:

\begin{itemize}

\item \texttt{X\--Stre\-am}~\cite{Roy:2013:XEG:2517349.2522740}. 
A system that provided an alternative view to the traditional \textit{vertex-centric} approach.
It is based on considering computation from the perspective of edges instead of vertices and experiments optimized the use of storage I/O both locally and on the cloud.
\texttt{X\--Stre\-am} is an open-source system written in \texttt{C++} which introduced the concept of \textit{edge\--cen\-tric} graph processing via streaming partitions.
\texttt{X\--Stre\-am} exposes an edge-centric sca\-tter\--ga\-ther programming model that was motivated by the lack of access locality when traversing edges, which makes it difficult to obtain good performance. 
State is maintained in vertices.
This tool uses the streaming partition, which works well with RAM and secondary (SSD and Magnetic Disk) storage types.
It does not provide any way by which to iterate over the edges or updates of a vertex.
A sequential access to vertices leads to random access of edges which decreases performance.
\texttt{X\--Stre\-am} is innovative in the sense that it enforces sequential processing of edges (edge-centric) in order to improve performance.
It is available~\cite{xstream_github} under the \texttt{A\-pa\-che} \texttt{Li\-cen\-se} \texttt{2.0}.

\item \texttt{Cha\-os}~\cite{Roy:2015:CSG:2815400.2815408}. 
A system written in \texttt{C++} which had its foundations on \texttt{X\-Stream}.
On top of the secondary storage studies performed in the past, graph processing in \texttt{Cha\-os} achieves scalability with multiple machines in a cluster computing system.
It is based on different functionalities: load balancing, randomized work stealing, sequential access to storage and an adaptation of \texttt{X\--Stre\-am}'s streaming partitions to enable parallel execution.
\texttt{Cha\-os} is composed of a storage sub-system and a computation sub-system.
The former exists concretely as a storage engine in each machine.
Its concern is that of providing edges, vertices and updates to the computation sub-system.
Previous work on \texttt{X\--Stre\-am} highlighted that the primary resource bottleneck is the storage device bandwidth.
In \texttt{Cha\-os}, the storage and computation engines' communication is designed in a way that storage devices are busy all the time -- thus optimizing for the bandwidth bottleneck.
It was released~\cite{chaos_github} under the \texttt{A\-pa\-che} \texttt{Li\-cen\-se} \texttt{2.0}.

\end{itemize}

The following graph processing systems were grou\-ped together because each of the improvements they proposed are important concerns to be aware of in designing graph processing systems. 

\begin{itemize}

\item \texttt{Pow\-er\-Ly\-ra}~\cite{chen2015powerlyra} is a graph computation engine written in \texttt{C++} which adopts different partitioning and computing strategies depending on vertex types. 
The authors note that most systems use a \textit{one-size-fits-all} approach.
They note that \texttt{Pre\-gel} and \texttt{Gra\-ph\-Lab} focus in hiding latency by evenly distributing vertices to machines, making resources locally accessible.
This may result in imbalanced computation and communication for vertices with higher degrees (frequent in scale-free graphs).
Another provided design example is that of \texttt{Po\-wer\-Gra\-ph} and \texttt{GraphX} which focus on evenly parallelizing the computation by partitioning edges among machines, incurring communication cos\-ts on vertices, even those with low degrees. 
\texttt{Pow\-er\-Ly\-ra} was released under the \texttt{A\-pa\-che} \texttt{Li\-cen\-se} \texttt{2.0}~\cite{powerlyra_github}.

\item \texttt{Ki\-neo\-gra\-ph}~\cite{Cheng:2012:KTP:2168836.2168846} is a system which combines snapshots allowing full processing in the background and explicit alternative/custom functions that, besides assessing updates' impact, also apply them incrementally, pro\-pa\-ga\-ting th\-eir outcome across the gra\-ph. 
It is a distributed system to capture the relations in incoming data feeds, built to maintain timely updates against a continuous flow of new data.
Its architecture uses \textit{ingest} nodes to register graph update operations as identifiable transactions, which are then distribu\-ted to \textit{graph} nodes.
No\-des of the latter type form a distributed in-memory key/value store.
\texttt{Ki\-neo\-gra\-ph} performs computation on static snapshots, simplifying algorithm design. 
We did not find its source code online.

\item \texttt{Tor\-na\-do}~\cite{Shi:2016:TSR:2882903.2882950} is a system for real-time iterative analysis over evolving data.
It was implemented over \texttt{Apache Storm} and provides an asynchronous boun\-ded iteration model, offering fine-grained updates while ensuring correctness. 
It is based on the observations that: \textit{1)} loops starting from \textit{good enough} guesses usually converge quickly; \textit{2)} for many iterative methods, the running time is closely related to the approximation error.
From this, an execution model was built where a main loop continuously gathers incoming data and instantly approximates the results.
Whenever a result request is received, the model creates a branch loop from the main loop.
This branch loop uses the most recent approximations as a guess for the algorithm. 
We did not find its source code online.

\item \texttt{Ki\-ck\-Star\-ter}~\cite{Vora:2017:KFA:3037697.3037748} is a system that debuted a runtime technique for trimming approximation values for subsets of vertices impacted by edge deletions.
The removal of edges may invalidate the convergence of approximate values pertaining monotonic algorithms.
\texttt{Kick\-Starter} deals with this by identifying values impacted by edge deletions and adapting the network impacts before the following computation, achieving good results on real-world use-cases.
Despite this, by focusing on monotonic graph algorithms, its scope is narrowed to selection-based algorithms. 
For this class, updating a vertex value implies choosing a nei\-gh\-bour under some criteria.
\texttt{Ki\-ck\-Star\-ter} is now known as \texttt{GraphBolt}, a recent work~\cite{graphbolt_github,Mariappan:2019:GDS:3302424.3303974} licensed under the \texttt{MIT License}~\cite{graphbolt_github} whi\-ch offers a generalized incremental programming model enabling development of incremental versions of complex aggregations.
Both were written in \texttt{C++}.

\item \texttt{Pi\-xie}~\cite{Eksombatchai:2018:PSR:3178876.3186183} is a graph-based scalable real-time recommendation system used at Pinterest.  
Using a set of user-specific \textit{pins} (in Pinterest, users have boards in which they store pins, where each pin is a combination of image and text), \texttt{Pi\-xie} chooses in real-time the pins that are most related to the query, out of billions of candidates. 
With this system, Pinterest was able to execute a custom (Pixie Random Walk) algorithm on an object graph of 3 billion vertices and 17 billion edges. 
On a single server, they were able to serve around 1200 recommendation requests per seconds with 60 millisecond latency. 
The authors note that the deployment of \texttt{Pi\-xie} benefited from large RAM machines, using a cluster of Amazon AWS r3.8xlarge machines with 244GB RAM. 
They fitted the pruned Pinterest graph (3 billion vertices, 17 billion edges) in around 120GB of RAM, in a setup that yielded the following advantages: random walk not forced to cross machines, which increases performance; multiple walks can be executed on the graph in parallel; the system can be parallelized and scaled by adding more machines to the cluster.
This system is a relevant case study (of applying graph theory to recommendation systems at scale) as a scalable system for processing on large graphs a biased random walk algorithm (with user-specific preferences) while using graph pruning techniques to disregard large boards that are too diverse and diffuse the random walk (the non-pruned graph has 7 billion vertices and 100 bilion edges). 
We did not find the source code available online.

\item \texttt{Flow\-Gra\-ph}~\cite{chaudhry2019flowgraph} is a system that proposes a syntax for a language to detect temporal patterns in large-scale graphs and introduces a novel data structure to efficiently store results of graph computations. 
This system is a unification of graph data with stream processing considering the changes of the graph as a stream to be processed and offering an API to satisfy temporal patterns.
We did not find its source code available. 

\item \texttt{GPS}~\cite{salihoglu2013gps} is an open-source (\texttt{BSD License}) scalable graph processing system written in \texttt{Ja\-va} and allowing fault-tolerant and ea\-sy\--to\--pro\-gram algorithm execution on very large graphs. 
It adopts \texttt{Pre\-gel}'s vertex-centric API and extends it with: features to make global computations easier to express and more efficient; dynamic repartitioning scheme to reassign vertices to different workers during computation based on messaging patterns; distribution of high-degree vertex adjacency lists across all computer nodes to improve performance (something that \texttt{PowerGraph} and \texttt{Pow\-er\-Ly\-ra} later adopted). 
It was designed to run on a cluster of machines such as Amazon EC2, over which the authors tested their work.
\texttt{GPS}'s initial version was run on up to 100 Amazon EC2 large instances and on graphs of up to 250 million vertices and 10 billion edges. 
It is open-source and available under the \texttt{BSD License}~\cite{gps_source}.

\item \texttt{Go\-F\-Fis\-h}~\cite{simmhan2014goffish} is a sub-graph centric programming abstraction and framework co-designed with a distributed persistent graph storage for large scale graph analytics on commodity clusters, aiming to combine the scalability of the vertex-centric (TLAV) approach with flexibility of shared-memory sub-graph computation (TLAG).  
It is written in \texttt{Ja\-va}. 
\texttt{Go\-F\-Fis\-h} states that two sub-graphs many not share the same vertex, but they can have remote edges that connect their vertices, provided that the sub-graphs are on different partitions. 
If two sub-graphs in the same partition share an edge, by definition they are merged into a single-sub-graph. 
It was evaluated with a cluster of 12 nodes each with 8-core Intel Xeon CPUs, 16 GB RAM and 1 TB SATA HDD.
The authors compare the execution of \texttt{Go\-F\-Fis\-h} (one worker per node) with \texttt{Gi\-ra\-ph} (default two workers per node), achieving faster execution times for algorithms such as PageRank, connected components and single-source shortest-paths.
Its source code is available though we did not find any information pertaining licensing. 
While its source code is available~\cite{goffish_github}, we did not find information regarding licensing.

\item \texttt{F\-B\-S\-Gra\-ph}~\cite{zhang2017fbsgraph} presents a forward and backward sweeping execution method to accelerate state propagation for asynchronous graph processing. 
In asynchronous graph processing, each vertex maintains a state which can be asynchronously updated in an iterative fashion.
The method presented in \texttt{F\-B\-S\-Gra\-ph} relies on the observation that state can be propagated faster by processing vertices sequentially along the graph path in each round. 
They achieve greater execution speed when analyzing several graph algorithms across a set of datasets, comparing against systems such as \texttt{PowerGraph} and \texttt{Gra\-ph\-Lab}. 
We did not find its source available.

\item \texttt{Gra\-pH}~\cite{mayer2018graph} is a graph processing system written in \texttt{Ja\-va} that uses vertex-cut graph partitioning that takes into consideration the diversity of vertex traffic and the heterogeneous costs of the network. 
It relies on a strategy of adaptive edge migration to reduce the frequency of communication across expensive network links. 
For this work, the authors focused on vertex-cut as it has better partitioning properties for real-world graphs that have power-law degree distributions. 
\texttt{Gra\-pH} has two partitioning techniques, \textit{H-load} which is used for the initial partitioning of the graph so that each cluster worker can load it into local memory, and \textit{H-adapt}, a distributed edge migration algorithm to address the dynamic heterogeneity-aware partitioning problem. 
In evaluation, \texttt{Gra\-pH} outperformed \texttt{PowerGraph}'s vertex-cut partitioning algorithm with respect to communication costs.
While its source code is available~\cite{grapH_github}, we found no information on licensing.

\item \texttt{Ju\-lie\-n\-ne}~\cite{dhulipala2017julienne} is built over \texttt{Li\-gra} (\texttt{C++}) and provides an interface to maintain a collection of buckets under vertex insertions and bucket deletions. 
They evaluated under bucketing algorithms such as weighted breadth-first search, $k$-core and approximate set cover. 
The authors describe as \textit{bucketing-based} algorithms those that maintain vertices in a set of unordered buckets - and in each round, the algorithm extracts the vertices contained in the lowest (or highest) bucket to perform computation on them.
Then, it can update the buckets containing the extracted vertices or their neighbours. 
For example, weighted breadth-first search processes vertices level by level, where level $i$ contains all vertices at distance $i$ from the source vertex. 
The system was tested in a multi-core machine with 72 cores (4 CPUs at 2.4GHz) and 1TB of main memory, achieving performance improvements on several data sets when compared to systems such as \texttt{Galois}, base \texttt{Li\-gra} and \texttt{Galois}. 
We did not find its source code available.

\item \texttt{Gra\-ph\-D}~\cite{yan2017graphd} is an out-of-core system inspired by \texttt{Pre\-gel} and targeting efficient big graph processing using a small cluster of commodity machines connected by Gigabit Ethernet, contrasting with other out-of-core works that focus on specialized hardware. 
The authors focus on a setting that sees vertex-centric programs being data-intensive, as the CPU cost of computing a message is small when compared to the network transmission cost. 
\texttt{Gra\-ph\-D} masks disk I/O overhead with message transmission though parallelism of computation and communication. 
It eliminates the need for (expensive) external-memory join or group-by operations, which are required in other systems such as \texttt{Cha\-os}. 
It was evaluated on PageRank, single-source shortest-paths and connected components. 
\texttt{Gra\-ph\-D} was evaluated against distributed out-of-core systems \texttt{Pre\-ge\-lix}, \texttt{Ha\-Lo\-op} and \texttt{Cha\-os}, against single-machine systems \texttt{Gra\-ph\-Chi} and \texttt{X\--Stre\-am} and representative in-memory systems \texttt{Pre\-gel} and \texttt{Gi\-ra\-ph}, achieving better performance in some scenarios. 
We did not find its source available.

\item \texttt{Tur\-bo\-Gra\-ph++}~\cite{ko2018turbograph} is a graph analytics system that exploits external memory for scale-up without compromising efficiency. 
It introduced an abstraction called nested windowed streaming to achieve scalability and increase efficiency in processing neighbourhood-centric analytics (in which the total size of neighbourhoods around vertices can exceed the available memory budget). 
This streaming model regards a sequence of vertex values and an adjacency list stream.  
The goal is to efficiently support the $k$-walk neighbourhood query (a class of graph queries defined by the authors, where walks are enumerated and then computation is done for each one) with fixed size memory.
In the model, during user computation, they define an update stream as the sequence of updates generated to the ending vertex of each walk, with each update represented as a pair of target vertex ID and update value. 
\texttt{Tur\-bo\-Gra\-ph++} has the goal of balancing the workloads across machines, which requires balancing the number of edges and the number of high-degree and low-degree vertices among machines. 
It also focuses on balancing the number of vertices on each machine so that each one requires the same memory budget.
We did not find its source code available online.

\item \texttt{Gra\-ph\-In}~\cite{sengupta2016graphin} is a dynamic graph analytics framework proposed to handle the scale and evolution of real-world graphs. 
It aimed to improve over approaches to processing dynamic graphs which repeatedly run static graph analytics on stored snapshots. 
\texttt{Gra\-ph\-In} proposes an adaptation of gather-apply-scatter (GAS) called I-GAS which enables the implementation of incremental graph processing algorithms across multiple CPU cores. 
It also introduces an optimization heuristic to choose between static or dynamic execution based on built-in and user-defined graph properties.
Native and benchmarking code were implemented in \texttt{C++} and for experimental evaluation it was compared to \texttt{Gra\-ph\-Mat} and \texttt{STINGER}. 
The heuristic-base computation made \texttt{Gra\-ph\-In} faster than systems using fixed strategies.
We did not find its source code available.

\end{itemize}

The following works focus on specific techniques such as using specific hardware such as SSDs or GPUs. 
We first list frameworks and systems that were proposed in the last years to use the single-instruction multiple-data (SI\-MD) capabilities of G\-P\-Us for graph processing:
 
\begin{itemize}

\item \texttt{Map\-Gra\-ph}~\cite{fu2014mapgraph} is a high-performance parallel graph programming framework, able to achieve up to 3 billion traversed edges per second using a GPU.
It represents the graph with a compressed sparse row (CSR) data structure and chooses different scheduling strategies depending on the size of the \textit{frontier} (the set of vertices that are active in a given iteration). 
It encapsulates the complexity of the GPU architecture while enabling dynamic runtime decisions among several optimization strategies.
Users need only to write sequential \texttt{C++} code to use the framework.
The early \texttt{Map\-Gra\-ph} work was first available as an open-source project~\cite{mapgraph_github} licensed under the \texttt{A\-pa\-che} \texttt{Li\-cen\-se} \texttt{2.0}, but it has been integrated in the former line of products of \texttt{Blazegraph}, also available online\-\cite{blazegraph}.

\item \texttt{Cu\-Sha}~\cite{khorasani2014cusha} is a CUDA-based graph processing framework written in \texttt{C++} which was motivated by the negative impact that irregular memory accesses have on the compressed sparse row graph (CSR) representation. 
\texttt{Cu\-Sha} overcomes this by: \textit{1)} organizing the graph into autonomous sets of ordered edges called \textit{shards} (a representation they call \textit{G-Shards}) unto which GPU hardware resources are mapped for fully coalesced memory accesses; \textit{2)} accounting for input graph properties such as sparsity (the sparser the graph, the smaller the computation windows) to avoid GPU under-utilization (\textit{Concatenated Windows}, or \textit{CW}).
This framework allows users to define vertex-centric algorithms to process large graphs on GPU. 
It is open-source~\cite{cusha_github} and available under the \texttt{MIT License}.

\item \texttt{Gun\-ro\-ck}~\cite{wang2016gunrock,wang2017gunrock} is an open-source~\cite{gunrock_github} (\texttt{A\-pa\-che} \texttt{Li\-cen\-se} \texttt{2.0}) CUDA library for graph processing targeting the GPU and written in \texttt{C}. 
It implements a data-centric abstraction focused on operations on a vertex or edge frontier. 
For different graph algorithms, it achieved at least an order of magnitude speedup over \texttt{PowerGraph} and better performance than any other high-level GPU graph library at the time. 
Its operations are bulk-synchronous and manipulate a frontier, which is a subset of the edges or vertices within the graph that is relevant at a given moment in the computation. 
\texttt{Gun\-ro\-ck} couples high-performance GPU computing primitives and optimization strategies with a high-level programming model to quickly develop new graph primitives. 
It was evaluated using brea\-dth\--fir\-st sear\-ch, de\-pth\--fir\-st sear\-ch, single-source shortest paths, connected components and PageRank. 

\item \texttt{Lux}~\cite{jia2017distributed} is a distributed multi-GPU system written in \texttt{C++} for fast graph processing by exploiting aggregate memory bandwidth of multiple GPUs and the locality of the memory hierarchy of multi-GPU clusters. 
It proposes a dynamic graph repartitioning strategy to enable well-balanced distribution of workload with minimal overhead (improving performance by up to 50\%), as well as a performance model providing insight on how to choose the optimal number of nodes and GPUs to optimize performance. 
\texttt{Lux} is aimed at graph programs that can be written with iterative computations. 
Vertex properties are read-only in each iteration, with updates becoming visible at the end of an iteration. 
It offers two execution models: \textit{pull} which optimizes run-time performance of GPUs (enables optimizations like caching and locally aggregating updates in GPU shared me\-mory); and \textit{push}, which optimizes algorithmic efficiency (maintains a frontier queue and only performs computation over the out-edges of vertices in the frontier). 
Its source code is available~\cite{lux_github} under \texttt{A\-pa\-che} \texttt{Li\-cen\-se} \texttt{2.0}.

\item \texttt{Frog}~\cite{shi2017frog} is a light-weight asynchronous processing framework written in \texttt{C}. 
The authors note that common coloring algorithms may suffer from low parallelism due to a large number of colors being needed to process large graphs with billions of vertices. 
\texttt{Frog} separates vertex processing based on color distribution. 
They propose an efficient hybrid graph coloring algorithm, relying on a relaxed pre-partition method to solve vertex classification with a lower number of colors, without forcing all adjacent vertices to be assigned different colors.  
The execution engine of \texttt{Frog} scans the graph by color, and all vertices under the same color are updated in parallel in the GPU. 
For large graphs, when processing each partition, the data transfers are overlapped with GPU kernel function executions, minimizing PCIe data transfer overhead. 
It is open-source~\cite{frog_github} and licensed under the \texttt{GNU General Public License 2.0}.

\item \texttt{Aspen}~\cite{dhulipala2019low} is a graph-streaming extension of the \texttt{Li\-gra} interface, supporting graph updates. 
To support this, the authors developed and presented the $C$-tree data structure which achie\-ves good cache locality, lowers space use and has operations which are efficient from a theoretical perspective.
It applies a chunking sche\-me over the tree, storing multiple elements in a tree-node. 
The scheme takes the ordered set of elements that are represented. 
More relevant elements are stored in tree nodes, while the remaining ones are associated in tails of the tree nodes. 
It employs compression and supports parallelism.
The authors evaluate it with the largest publicly-available graph, which has more than two hundred billion edges on a multi-core server with 1 TB memory.
Source code is available online~\cite{aspen_github} albeit no license information was provided.

\item \texttt{Glu\-on}~\cite{dathathri2018gluon} was introduced as a new approach to create distributed-memory graph analytics systems able to use heterogeneity in partitioning policies, processor types (GPU and CPU) and programming models.
To use \texttt{Glu\-on}, programmers implement applications in shared-memory programming systems of their choice and then interface the applications with \texttt{Glu\-on} to enable execution on heterogeneous clusters. 
\texttt{Glu\-on} optimizes communication by taking advantage of temporal and structural invariants of graph partitioning policies.  
It runs on shared-memory NU\-MA platforms and NVI\-DIA GPUs. 
Its programming model offers a small number of programming patterns implemented in \texttt{C++}, its library offers concurrent data structures, schedulers and memory allocators and the runtime executes programs in parallel, using parallelization strategies as optimistic and round-based execution.
\texttt{Glu\-on} is available~\cite{gluon_github} under the \texttt{3\--Clau\-se} \texttt{B\-S\-D} \texttt{Li\-cen\-se}.

\item \texttt{Hornet}~\cite{busato2018hornet} is a data structure for efficient computation of dynamic sparse graphs and matrices using GPUs.
It is platform-independent and implements its own memory allocation operation instead of standard function calls. 
The implementation uses an internal data manager which makes use of block arrays to store adjacency lists, a bit tree for finding and reclaiming empty memory blocks and $B^{+}$ trees to manage them. 
It was evaluated using an NVIDIA Tesla GPU and experiments targeted the update rates it supports, algorithms such as brea\-dth\--fir\-st search (BFS) and sparse matrix-vector multiplication.
\texttt{Hornet} is available~\cite{hornet_github} under the \texttt{3\--Clau\-se} \texttt{B\-S\-D} \texttt{Li\-cen\-se}.

\item \texttt{faimGraph}~\cite{winter2018faimgraph} introduced a fully-dynamic graph data structure performing autonomous memory management on the GPU. 
It enables complete reuse of memory and reduces memory requirements and fragmentation. 
The implementation has a vertex-centric update scheme that allows for edge updating in a lock-free way. 
It reuses free vertex indices to achieve efficient vertex insertion and deletion, and does not require restarting as a result of a large number of edge updates.
\texttt{faimGraph} was benchmarked against \texttt{Hornet} on an NVIDIA Geforce GTX Titan Xp GPU using algorithms such as PageRank and triangle counting.
Source code is available online~\cite{faimgraph_github} without a specified license.

\item \texttt{Gra\-ph\-Ca\-ge}~\cite{chen2019graphcage} is a cache-centric optimization framework to enable highly efficient graph processing on GPUs. 
It was motivated by the random memory accesses which are generated by sparse graph data structures, which increase memory access latency.
The authors note that conventional cache-blocking suffers from re\-pea\-ted accesses when processing large graphs on GPUs, and propose a throughput-oriented ca\-che blocking scheme (\textit{TO\-CAB}). 
\texttt{Gra\-ph\-Ca\-ge} applies the scheme to both push and pull directions and coordinates with load balancing strategies by considering sparsity of sub-graphs.
This technique is applied to traversal-based algorithms by considering the benefit and overhead in different iterations with working sets of different sizes.
In its evaluation, \texttt{Gra\-ph\-Ca\-ge} achieved in average lower execution times for one PageRank iteration compared to both \texttt{Gun\-ro\-ck} and \texttt{Cu\-Sha}. 
We did not find its source code available.

\end{itemize}

For more information on GPU use cases for graph processing approaches, we point the readers to~\cite{shi2018graph}.

\begin{itemize}

\item \texttt{Fla\-sh\-Gra\-ph}~\cite{zheng2015flashgraph} is a graph processing engine implemented in \texttt{C++} over a user-space SSD file system designed for high IOS and very high levels of parallelism. 
Vertex state is stored in memory while edge lists are on SSDs. 
Latency is hidden by overlapping computation with I/O, a concept similar to \texttt{X\--Stre\-am} and \texttt{Cha\-os}, and edges lists are only accessed if requested by applications from SSDs. 
\texttt{Fla\-sh\-Gra\-ph} has a vertex-centric (TLAV) interface, its designed to reduce CPU overhead and increase throughput by conservatively merging I/O requests, and the authors demonstrate that \texttt{Fla\-sh\-Gra\-ph} in semi-external memory executes many algorithms with a performance of up to 80\% of the in-memory implementation and It also outperformed \texttt{PowerGraph}. 
It is open-source~\cite{flashgraph_github} under the \texttt{A\-pa\-che} \texttt{Li\-cen\-se} \texttt{2.0}.

\item \texttt{Gra\-ph\-S\-S\-D}~\cite{matam2019graphssd} is a semantic-aware SSD framework and full system solution to store, access and execute graph analytics. 
Instead of considering storage as a set of blocks, it accounts for graph structure while choosing graph layout, access and update mechanisms. 
\texttt{Gra\-ph\-S\-S\-D} innovates by considering a vertex-to-page mapping scheme and uses advanced knowledge of flash properties to reduce page accesses. 
It offers a simple API to ease development of applications accessing graphs as native data and its evaluation showcased average performance gains for basic graph data fetch functions on breadth-first search, connected components, random-walk, maximal independent set and PageRank. 
We did not find its source available.

\end{itemize}

In Table~\ref{table:graph_system_features} we summarize distinguishing features and licenses for the graph processing systems detailed in this section. 
The last reference in front of every system name is its open-source code repository, when available.
The second group from the top (\texttt{P\-B\-G\-L}, \texttt{Com\-b\-BLAS} and \texttt{Ha\-vo\-q\-GT}) contains systems which use multiple machines for computation but not in the typical cluster scenario. 
Instead, they are characterized by using specific machines for high-performance computing.

\begin{table*}[h!] 
	\scriptsize
	\centering
	\setlength\tabcolsep{1.5pt} 
	\begin{tabularx}{10.0cm}{p{2.8cm} |p{0.166cm}| p{0.166cm}| p{0.166cm}| c c c  } 
		\toprule 
		
		\emph{System} & \rotatebox[origin=c]{90}{\emph{Multi-core}}& \rotatebox[origin=c]{90}{\emph{GPU}} & \rotatebox[origin=c]{90}{\emph{Cluster}}  & \emph{Languages}  &  \emph{License} & \emph{Notes} \\\midrule
		
		\texttt{Gra\-ph\-Lab}~\cite{Low_graphlab:a,graphlab_github}    & \blackcirclemarker  &   & \blackcirclemarker  & \texttt{C++} & \texttt{AL 2.0} & \textit{N/A} \\ 
		\texttt{GRA\-CE}~\cite{wang2013asynchronous}    & \blackcirclemarker  &  &  \blackcirclemarker  &  \texttt{C++} & \textit{Unavailable} & \textit{N/A}  \\ 
		\texttt{Li\-gra}~\cite{shun2013ligra,ligra_github} & \blackcirclemarker  &   & &  \texttt{C++}  & \texttt{MIT} & \textit{N/A} \\ 
		\texttt{Rin\-go}~\cite{Perez:2015:RIG:2723372.2735369,ringo_github} & \blackcirclemarker  &  &  &  \texttt{C++}, \texttt{Py\-thon} & \texttt{BSD} & \textit{N/A} \\
		\texttt{Po\-ly\-mer}~\cite{zhang2015numa,polymer_github} & \blackcirclemarker   &   &    &  \texttt{C++}&  \texttt{AL 2.0}  & \textit{N/A} \\ 
		\texttt{Gra\-ph\-Mat}~\cite{sundaram2015graphmat,graphmat_github} & \blackcirclemarker  &   &   &  \texttt{C++}  & \textit{Custom} & \textit{N/A} \\ 
		\texttt{Mo\-saic}~\cite{Maass:2017:MPT:3064176.3064191,mosaic_github} & \blackcirclemarker  &   &   &  \texttt{C++}  & \texttt{MIT} & \textit{Fast storage} \\
		
		\midrule 
		
		\texttt{P\-B\-G\-L}~\cite{gregor2005parallel,PBGL_github} & \blackcirclemarker  &   &  & \texttt{C++}  & \textit{Custom} & \textit{Hardware} \\ 
		\texttt{Com\-b\-BLAS}~\cite{bulucc2011combinatorial,comblas_repo} & \blackcirclemarker &  &   & \texttt{C++} & \textit{Custom} & \textit{Hardware}\\ 
		\texttt{Ha\-vo\-q\-GT}~\cite{pearce2013scaling,havocgt_repo} & \blackcirclemarker  &   &  & \texttt{C++} & \texttt{GNU LGPL 2.1} & \textit{Hardware} \\ 

		\midrule 

		\texttt{A\-pa\-che} \texttt{Gi\-ra\-ph}~\cite{ching2013scaling,giraph_github} & \blackcirclemarker  &   &  \blackcirclemarker & \texttt{Ja\-va} & \texttt{AL 2.0} & \textit{N/A} \\ 
		\texttt{Nai\-ad}~\cite{Murray:2013:NTD:2517349.2522738,naiad_github} & \blackcirclemarker  &  &  & \texttt{C\#} & \texttt{AL 2.0} & \textit{N/A} \\ 
		\texttt{A\-pa\-che} \texttt{Flink}~\cite{carbone2015apache,flink_github} & \blackcirclemarker  &  & \blackcirclemarker   & \texttt{Ja\-va}, \texttt{Py\-thon}, \texttt{Scala}  &  \texttt{AL 2.0}  & \textit{N/A} \\
		\texttt{A\-pa\-che} \texttt{Spark}~\cite{Zaharia:2010:SCC:1863103.1863113,spark_github}    & \blackcirclemarker  &  & \blackcirclemarker & \texttt{Ja\-va}, \texttt{Py\-thon}, \texttt{Scala}  & \texttt{AL 2.0}   &  \textit{N/A}   \\ 
		\texttt{Gra\-ph\-Tau}~\cite{iyer2016time}    & \blackcirclemarker  &  & \blackcirclemarker  & \texttt{Ja\-va}, \texttt{Scala}  & \textit{Unavailable}  &  \textit{N/A}    \\ 
		\texttt{Tink}~\cite{lightenberg2018tink,tink_github}    & \blackcirclemarker  &  & \blackcirclemarker & \texttt{Ja\-va}, \texttt{Scala} & \texttt{AL 2.0}   & \textit{N/A}    \\

		\midrule 

		\texttt{X\--Stre\-am}~\cite{Roy:2013:XEG:2517349.2522740,xstream_github}    & \blackcirclemarker  &  &  & \texttt{C++} &  \texttt{AL 2.0} &   \textit{N/A} \\ 
		\texttt{Cha\-os}~\cite{Roy:2015:CSG:2815400.2815408,chaos_github}   & \blackcirclemarker  &  & \blackcirclemarker & \texttt{C++} & \texttt{AL 2.0} & \textit{N/A}   \\ 
		
		\midrule 
		
		\texttt{Pow\-er\-Ly\-ra}~\cite{chen2015powerlyra,powerlyra_github} & \blackcirclemarker  &  & \blackcirclemarker & \texttt{C++} & \texttt{AL 2.0} & \textit{N/A}  \\
		\texttt{Ki\-neo\-gra\-ph}~\cite{Cheng:2012:KTP:2168836.2168846,graphbolt_github} & \blackcirclemarker  &  & \blackcirclemarker  & \textit{Unknown} & \textit{Unavailable} & \textit{N/A}  \\
		\texttt{Tor\-na\-do}~\cite{Shi:2016:TSR:2882903.2882950} & \blackcirclemarker  &  & \blackcirclemarker   & \textit{Unknown} & \textit{Unavailable} & \textit{N/A}  \\
		\texttt{Ki\-ck\-Star\-ter}~\cite{Vora:2017:KFA:3037697.3037748} & \blackcirclemarker  &  & \blackcirclemarker  & \texttt{C++} & \texttt{MIT} & \textit{N/A}  \\
		\texttt{Pi\-xie}~\cite{Eksombatchai:2018:PSR:3178876.3186183} & \blackcirclemarker  &  & \blackcirclemarker & \textit{Unknown} & \textit{Unavailable} & \textit{N/A}  \\
		\texttt{Flow\-Gra\-ph}~\cite{chaudhry2019flowgraph} & \blackcirclemarker  &  & \blackcirclemarker   & \textit{Unknown} & \textit{Unavailable} & \textit{N/A}  \\
		\texttt{GPS}~\cite{salihoglu2013gps,gps_source} & \blackcirclemarker  &  & \blackcirclemarker & \texttt{Ja\-va} & \texttt{BSD} & \textit{N/A}  \\
		
		\texttt{Go\-F\-Fis\-h}~\cite{simmhan2014goffish,goffish_github} & \blackcirclemarker  &  & \blackcirclemarker  & \texttt{Ja\-va}  & \textit{Unknown} & \textit{Copyright}  \\
		\texttt{F\-B\-S\-Gra\-ph}~\cite{zhang2017fbsgraph} & \blackcirclemarker  & & \blackcirclemarker    & \texttt{Unknown} & \textit{Unavailable} & \textit{N/A}  \\
		\texttt{Gra\-pH}~\cite{mayer2018graph,grapH_github} & \blackcirclemarker  &  & \blackcirclemarker  & \texttt{Ja\-va} & \textit{Unknown} & \textit{Copyright}  \\
		
		\texttt{Ju\-lie\-n\-ne}~\cite{dhulipala2017julienne} & \blackcirclemarker  &  &   & \texttt{C++} & \textit{Unavailable} & \textit{N/A}  \\
		\texttt{Gra\-ph\-D}~\cite{yan2017graphd} & \blackcirclemarker  &  & \blackcirclemarker   & \texttt{Unknown} & \textit{Unavailable} & \textit{N/A}  \\
		\texttt{Tur\-bo\-Gra\-ph++}~\cite{ko2018turbograph} & \blackcirclemarker  &  & \blackcirclemarker  & \texttt{Unknown} & \textit{Unavailable} & \textit{N/A}  \\
		\texttt{Gra\-ph\-In}~\cite{sengupta2016graphin} & \blackcirclemarker  &  &  & \texttt{C++}  & \textit{Unavailable} & \textit{N/A}  \\

		\midrule 

		\texttt{Map\-Gra\-ph}~\cite{fu2014mapgraph,mapgraph_github}    &  & \blackcirclemarker  &   & \texttt{C++} & \texttt{AL 2.0} & \textit{Discontinued} \\ 
		\texttt{Cu\-Sha}~\cite{khorasani2014cusha,cusha_github}    &   & \blackcirclemarker &   & \texttt{C++} & \texttt{MIT} & \textit{N/A}  \\ 
		\texttt{Gun\-ro\-ck}~\cite{wang2016gunrock,wang2017gunrock,gunrock_github}    &   & \blackcirclemarker &   & \texttt{C} &  \texttt{AL 2.0} & \textit{N/A}  \\ 
		\texttt{Lux}~\cite{jia2017distributed,lux_github}    & \blackcirclemarker  & \blackcirclemarker  & \blackcirclemarker & \texttt{C++}  & \texttt{AL 2.0} & \textit{N/A}  \\ 
		\texttt{Frog}~\cite{shi2017frog,frog_github}    &   & \blackcirclemarker &  & \texttt{C} & \texttt{GPL 2.0} & \textit{N/A}  \\ 
		\texttt{Glu\-on}~\cite{dathathri2018gluon,gluon_github}    & \blackcirclemarker  & \blackcirclemarker &   & \texttt{C++}  & \textit{3C BSD} & \textit{N/A} \\
		\texttt{Gra\-ph\-Ca\-ge}~\cite{chen2019graphcage}    &   & \blackcirclemarker  &   & \texttt{Unknown} & \textit{Unavailable} & \textit{N/A}  \\
		
		\midrule 
		
		\texttt{Fla\-sh\-Gra\-ph}~\cite{zheng2015flashgraph,flashgraph_github}    &  &  &    & \texttt{C++}  & \texttt{AL 2.0} & \textit{SSDs} \\
		\texttt{Gra\-ph\-S\-S\-D}~\cite{matam2019graphssd}    &   &  &   & \texttt{Unknown} & \textit{Unavailable} & \textit{SSDs}  \\

		\bottomrule 
	\end{tabularx}
	\caption{Summary of graph system distinctive features. Circle \blackcirclemarker\ on the \emph{Multi-core}, \emph{GPU} and \emph{Cluster} columns indicate that option is supported.  \emph{Languages} lists the programming languages the systems were written in. \emph{License} lists the licenses of the open-source project or of the free edition of a commercial product: \texttt{AL 2.0} is \texttt{A\-pa\-che} \texttt{Li\-cen\-se} \texttt{2.0}, \texttt{CC 1.0} is \texttt{Commons Clause 1.0}, \texttt{(GPL) v3} is \texttt{GNU General Public License (GPL) v3}. \emph{Notes} covers additional information, with \textit{Copyright} meaning that it may be illegal to reuse the source code.}
	\label{table:graph_system_features}
\end{table*}

\section{Graph Databases}\label{sec:graph-databases}

Here we list and describe different graph databases which we group together according to their type of supported graph model (e.g., property graph model, \texttt{RDF}, hybrid). 
Upon developing this list, we observe regarding graph query languages that they either implement their own custom language (neglecting interoperability even though there may be automatic converters developed by third parties), or they use those that have been improved, standardized and adopted by many projects (e.g., \texttt{Cy\-pher}, \texttt{Grem\-lin}, \texttt{SPAR\-QL}). 
The list includes both open-source and commercial products, as well as database systems implemented to explore novel techniques as part of research activities.
We showcase these different graph database solutions in Table~\ref{table:graph_database_features}, in which we present relevant properties for assessing them. 
We consider these properties to be outstanding features that are important for developers and users to assess their suitability for their use-cases.
They provide information on guarantees of graph data storage, parallelism of computation and supported graph sizes, visualization capabilities and the ability to use them freely (regarding both open-source as well as free versions of commercial products):

\begin{itemize}

\item Is it ACID compliant (\textit{ACID})?

\item Does the specific graph database ecosystem offer analytics capabilities (e.g., first-party, third-party, none) to complement the \texttt{OLTP} focus of the graph database itself  (\textit{Visual})?

\item Does it offer scale-out/horizontal scalability (e.g., Neo4j is very well-known but it does not support sharding, only replication when using multiple machines) to improve performance and support bigger graphs (\textit{Scale})?

\item What graph models does it support, is it for example \texttt{RDF}, property graph or others (\textit{Models})?

\item What languages can be used to program functionality that uses the database product, for example \texttt{Go}, \texttt{Ja\-va}, \texttt{.NET}, \texttt{Py\-thon} or others (\textit{Languages})?

\item What graph query languages (e.g., \texttt{Cy\-pher}, \texttt{Grem\-lin}, \texttt{SPAR\-QL}, custom) does it support (\textit{GQLs})?

\item What license (or types of licenses) is it subjected to - for example, is it a common license type or other (\textit{License})?

\item For database products that have both free and commercial/proprietary editions, what are the features missing from the free versions - the focus here is not on marketing-speech such as five-nines availability, but features that are easy to implement but are clearly missing from the free version to stimulate purchases (\textit{Limits})?

\end{itemize}

For further information on the evolution of graph databases, we point the reader to~\cite{angles2008survey,pokorny2015graph}, which covers, among other aspects, data models and query languages. 
The following graph databases we list are focused on the property graph model. 
Their information describes the use of this model; we found no graph query languages for \texttt{RDF} in their features, even if these databases are described as multi-model:

\begin{itemize}

\item \texttt{Alibaba Graph Database (GD)}~\cite{alibaba_graphdb} is a cloud-oriented (thus supporting horizontal scalability) graph database service supporting ACID transactions and the \texttt{Tin\-ker\-Pop} stack.
It supports the property graph model, the \texttt{Grem\-lin} query language and there are programming interfaces for \texttt{Go}, \texttt{Ja\-va}, \texttt{.NET}, \texttt{Node.js} and \texttt{Py\-thon}.
We did not find references to visualization capabilities (visual feedback to query construction and execution).

\item \texttt{Chro\-no\-Gra\-ph}~\cite{haeusler2017chronograph} is a \texttt{Tin\-ker\-Pop}-compliant (offering \texttt{Grem\-lin} to query the data in property graph model) graph database supporting ACID transactions, system-time content versioning and analysis. 
It is implemented as a key-value store enhanced with temporal information, using a B-tree data structure.
The project is available online~\cite{chronograph_github} under the \texttt{aGPL v3} (open-source and academic purposes) and commercial licenses are available on demand.
We did not find visualization functionalities accompanying it.

\item \texttt{DataStax Enterprise Graph (DSE)}~\cite{datastax} is a proprietary fork of \texttt{Titan}, licensed under the \texttt{A\-pa\-che} \texttt{Li\-cen\-se} \texttt{2.0} and supporting \texttt{Ja\-va}.
It integrates with the \texttt{Ca\-s\-san\-dra}~\cite{Lakshman:2010:CDS:1773912.1773922} distributed database (over which it provides graph data models) and supports \texttt{Tin\-ker\-Pop} (pro\-per\-ty gra\-ph with \texttt{Grem\-lin}).
It is complemented by the \texttt{DataStax Studio}, which allows for interactive querying and visualization of graph data similarly to \texttt{Neo4j} and \texttt{SAP Hana Graph}.

\item \texttt{D\-gra\-ph}~\cite{dgraph_graphql,dgraph_github} was written in \texttt{Go} and it is a distributed graph database offering horizontal scaling and ACID properties.
It is built to reduce disk seeks and minimize network usage footprint in cluster scenarios.
\texttt{D\-gra\-ph} is licensed under two licenses: the \texttt{A\-pa\-che} \texttt{Li\-cen\-se} \texttt{2.0} and a \texttt{Dgraph Community License}.
It automatically moves data to rebalance cluster shards.
It uses a simplified version of the \texttt{Gra\-ph\-QL} query language.
Support for \texttt{Grem\-lin} or \texttt{Cy\-pher} has been mentioned for the future but will depend on community efforts. 
\texttt{D\-gra\-ph} has a scalability advantage over \texttt{Neo4j} as the latter may have multiple servers but they are merely replicas, while the former can grow horizontally (vertical scaling is expensive).
There is a proprietary enterprise version (conditions specified under a custom \texttt{Dgraph Community License}) with advanced features for backups and encryption.
Official clients include \texttt{Ja\-va}, \texttt{Ja\-va\-Scri\-pt} and \texttt{Py\-thon}.
To the best of our knowledge \texttt{D\-gra\-ph} does not offer visualization and global analytics functionalities.

\item \texttt{Graphflow}~\cite{kankanamge2017graphflow,10.14778/3342263.3342643,mhedhbi2020a+} was released as a prototype active graph database.
It is an in-memory graph store supporting the property graph model and supports one-time as well as continuous sub-graph queries.
\texttt{Graphflow} supports this using a one-time query processor  called \textit{Generic Join} and a \textit{Delta Join} which enables the continuous sub-graph queries. 
It extends the \texttt{openCypher} language with triggers to perform actions upon certain conditions. 
We did not find information regarding its direct use beyond academic purposes nor about supporting ACID transactions.
Its code is available online~\cite{graphflow_github} under the \texttt{A\-pa\-che} \texttt{Li\-cen\-se} \texttt{2.0}.

\item \texttt{Ja\-nus\-Gra\-ph}~\cite{janus} is an open-source project licensed~\cite{janus_github} under the \texttt{A\-pa\-che} \texttt{Li\-cen\-se} \texttt{2.0}.
A database optimized for storing (in adjacency list format) and querying large graphs with (billions of) edges and vertices distributed a\-cro\-ss a multi-machine cluster with ACID transactions.
\texttt{Ja\-nus\-Gra\-ph}, which debuted in 2017, is based on the source \texttt{Ja\-va} code base of the \texttt{Titan} graph database project and is supported by the likes of Google, IBM and the Linux Foundation, to name a few.
Like \texttt{Titan}, it supports \texttt{Ca\-s\-san\-dra}, \texttt{HBase} and \texttt{Ber\-ke\-ley\-DB}.
It was designed with a focus on scalability and it is in fact a transactional database aimed at handling many concurrent users, complex traversals and analytic queries.
\texttt{Ja\-nus\-Gra\-ph} can integrate platforms such as \texttt{Spark}, \texttt{Gi\-ra\-ph} and \texttt{Hadoop}.
It also natively integrates with the \texttt{Tin\-ker\-Pop} graph stack, supporting \texttt{Grem\-lin} applications, the query language and its graph server, with graphs in the property graph model. 
Due to supporting \texttt{Tin\-ker\-Pop}, one may use one of its drivers to use \texttt{Grem\-lin} from \texttt{Elixir}, \texttt{Go}, \texttt{Ja\-va}, \texttt{.NET}, \texttt{PHP}, \texttt{Py\-thon}, \texttt{Ru\-by} and \texttt{Scala}.
It supports global analytics using \texttt{Spark} integration as well.

\item \texttt{Ne\-bu\-la} \texttt{Gra\-ph} is an open-source graph database (available online~\cite{nebulagraph_github}) licensed under \texttt{A\-pa\-che} \texttt{Li\-cen\-se} \texttt{2.0}, provides a custom \texttt{Ne\-bu\-la Gra\-ph Que\-ry Lan\-gua\-ge (n\-G\-Q\-L)} with syntax close to \texttt{SQL} and \texttt{Cy\-pher} support is planned.
It supports the property graph model, ACID transactions and is implemented with a separation of storage and computation, being able to scale horizontally.
It supports multiple storage engines like \texttt{HBase}~\cite{DBLP:books/daglib/0027893} (implementing the graph logic over these key-value stores) and \texttt{Ro\-cks\-DB}\-\cite{rocksdb_github} and has clients in \texttt{Go}, \texttt{Ja\-va} and \texttt{Py\-thon}.
It also has the complementing \texttt{Nebula Graph Studio} for interactive visual querying and analytics. 

\item \texttt{Neo4j}~\cite{Webber:2012:PIN:2384716.2384777} is a graph database with multiple editions~\cite{neo4j_github}: a community edition licensed under the free \texttt{GNU General Public License (GPL) v3}, a commercial one and also an advanced edition licensed under \texttt{AGPLv3}.
It supports different programming languages \texttt{C\-/C++}, \texttt{Clojure}, \texttt{Go}, \texttt{Haskell}, \texttt{Ja\-va}, \texttt{Ja\-va\-Scri\-pt}, \texttt{.N\-ET}, \texttt{Perl}, \texttt{PHP}, \texttt{Py\-thon}, \texttt{R} and \texttt{Ru\-by}. 
\texttt{Neo4j} is optimized for highly-connected data.
It relies on methods of data access for graphs without considering data locality. 
\texttt{Neo4j}'s graph processing consists of mostly random data access.
For large graphs which require out-of-memory processing, the major performance bottleneck becomes the random access to secondary storage.
The authors created a system which supports ACID transactions, high availability, with operations that modify data occurring wi\-thin tran\-sac\-tions to guarantee consistency.
It uses the query language \texttt{Cy\-pher} and data is stored on disk as fixed-size records in linked lists.
\texttt{Neo4j} has a library offering many different graph algorithms. 
As far as we know, \texttt{Neo4j}'s scale-out capabilities are only true for read operations.
All writes are directed to the \texttt{Neo4j} cluster master, an architecture which has its limitations. 
Among other uses, \texttt{Neo4j} has also been employed for building applications using the \texttt{GRANDstack} framework~\cite{grandstack_farmework}. 
\texttt{Neo4j} also has an interactive graph explorer to query and update specific elements of the graph.

\item \texttt{Re\-dis\-Gra\-ph}~\cite{cailliau2019redisgraph} is a property graph database which uses sparse matrices to represent a gra\-ph's adjacency matrix and uses linear algebra for graph queries.
It uses custom memory-efficient data structures stored in RAM, having on-disk persistence and tabular result sets.
Queries may be written in a subset of \texttt{Cy\-pher} and are internally translated into linear algebra expressions. 
It has a custom license and client libraries for \texttt{Re\-dis\-Gra\-ph} have been developed in \texttt{Elixir}, \texttt{Go}, \texttt{Ja\-va}, \texttt{Ja\-va\-Scri\-pt}, \texttt{PHP}, \texttt{Py\-thon}, \texttt{Ru\-by} and \texttt{Rust}, complementing existing accesses that \texttt{Redis} already supports. 
As far as we know, it only works in single-server mode (bounded by the machine's RAM) and it does not support ACID properties.
It has a Community Edition under a custom license.
source code available online~\cite{redis_github} under a custom license.

\item \texttt{SAP Hana Graph}~\cite{rudolf2013graph,hwang2018graph,hana_github} is a column-oriented, in-memory relational database management system. 
It performs different type of data analysis, among which graph data processing with the property graph model and ACID transactions.
This graph functionality includes interpretation of \texttt{Cy\-pher} and a visual graph manipulation tool.
Its graph processing capabilities have served use cases like fraud detection and route planning.
We did not find source code available online.

\item \texttt{Spark\-see}~\cite{martinez2011dex,sparksee_former_dex} (formerly \texttt{DEX}) is a property graph database offering ACID transactions and representing the graph using bitmap data structures with high compression rates (with each bitmap partitioned into chunks that fit disk pages).
The graphs in \texttt{Spark\-see} are labeled multigraphs and it has multiple licenses depending on the purpose, with free licenses for evaluation, research and development.
It offers APIs in \texttt{C++}, \texttt{.NET}, \texttt{Ja\-va}, \texttt{Objective-C}, \texttt{Py\-thon} and mobile devices.
We found no capabilities for data visualization in \texttt{Spark\-see}, though it is able to export data to formats supported by third-party software.

\item \texttt{Ti\-ger\-Gra\-ph}~\cite{deutsch2019tigergraph,tigergraph} is a commercial graph database (formerly \texttt{GraphSQL}) implemented in \texttt{C++} and comes in three versions: developer edition (supporting only single-machine, no distribution and is only for non-production, research or educational purposes), cloud edition (as a managed service) and enterprise edition (allowing for horizontal scalability - distributed graphs).
It supports ACID consistency, access through a \texttt{REST} API, has a custom \texttt{SQL}-like query language (\texttt{GSQL}) and features a graphical user interface named \texttt{GraphStudio} to perform interactive graph data analytics. 
The \texttt{Ti\-ger\-Gra\-ph} model was designed to support graph vertices, edges and their attributes to support an engine that performs massively-parallel processing to compute queries and analytics.
Each vertex and edge acts as both a unit of storage and computation, integrating and extending both TL\-AV and TL\-EV para\-dig\-ms.
It supports the property graph model plus extensions to enable the massively-parallel processing.
We did not find available source code.

\item \texttt{Wea\-ver}~\cite{DBLP:journals/pvldb/DubeyHES16} is an open-source~\cite{weaver_github} gra\-ph da\-ta\-ba\-se (custom permissive license) for efficient, transactional graph analytics.
It introduced the concept of refinable timestamps.
It is a mechanism to obtain a rough ordering of distributed operations if that is sufficient, but also fine-grained orderings when they become necessary.
It is capable of distributing a graph across multiple shards while supporting concurrency.
Refinable timestamps allow for the existence of a multi-version graph: write operations use their timestamps as a mark for vertices and edges.
This allows for the existence of consistent versions of the graph so that long-running analysis queries can operate on a consistent version of the graph, as well as historical queries.
\texttt{Wea\-ver} is written in \texttt{C++}, offering binding options for \texttt{Py\-thon}. 
We did not find any support for popular graph query languages.

\end{itemize}

The following graph databases we list are focused on the \texttt{RDF} data model and variations (including support for the property graph model by representing them as \texttt{RDF}~\cite{hartig2014reconciliation}):

\begin{itemize}

\item \texttt{Alle\-gro\-Gra\-ph}~\cite{allegrograph} is a proprietary commercial graph database with clients under \texttt{Eclipse Public License v1 (EPL v1)} which supports several programming languages (\texttt{C\#}, \texttt{C}, \texttt{Common Lisp}, \texttt{Clojure}, \texttt{Ja\-va}, \texttt{Perl}, \texttt{Py\-thon}, \texttt{Scala}) that was purpose-built for \texttt{RDF} (triple-store).
It supports an array of mechanisms to access the information it stores, namely reasoning wi\-th ontology (\texttt{RDFS++\- Rea\-so\-ning}), materialized rea\-so\-ning (generating new triples based on inference rules - \texttt{OWL2 RL Materialized Reasoner}, \texttt{SPAR\-QL} queries, \texttt{Prolog} and also low-level APIs.

\item \texttt{Bla\-ze\-Gra\-ph}~\cite{blazegraph} (formerly \texttt{Bigdata}) is an \texttt{RDF} database able to support up to billions of edges in a single machine and available under the \texttt{G\-N\-U} \texttt{Ge\-ne\-ral} \texttt{Pu\-blic} \texttt{Li\-cen\-se v2.0} supporting \texttt{S\-PAR\-QL} and the \texttt{Tin\-ker\-Pop Blue\-prin\-ts} A\-P\-I.
It has \texttt{.NET} and \texttt{Py\-thon} clients.
One of its associated internal projects is \texttt{blazegraph-gremlin}, which allows the storage of property graphs internally in \texttt{RDF} format, which can then be queried with \texttt{SPAR\-QL}.
It essentially has an alternative approach to \texttt{RDF} reification, giving labeled property graph capabilities to \texttt{RDF} graphs, with the ability to query the graphs in \texttt{Grem\-lin} as well.

\item \texttt{Bri\-ght\-star\-DB} is an open-source~\cite{brightstardb_github} multi-threaded multi-platform (including mobile) \texttt{.N\-ET} \texttt{RDF} store, supporting \texttt{SPAR\-QL} and binding of \texttt{RDF} resources to \texttt{.NET} dynamic objects (it has tools to use \texttt{.NET} interfaces and generate concrete classes to persist their data in \texttt{Bri\-ght\-star\-DB}). 
It is licensed under the \texttt{MIT License} and there is also an Enterprise version.
\texttt{Bri\-ght\-star\-DB} supports single-threaded writes and multi-threaded reads, with ACID transactions. 
It does not support horizontal scaling.

\item \texttt{Cray Graph Engine (CGE)}~\cite{rickett2018loading} is an \texttt{RDF} tri\-ple database offering the \texttt{SPAR\-QL} query language. 
As a commercial product, it was designed while considering different architectures of proprietary systems (containing the \texttt{Cray Aries interconnect}~\cite{alverson2012cray}) of the company behind \texttt{GCE}.
It offers APIs for \texttt{Ja\-va}, \texttt{Py\-thon} and \texttt{Spark}, having a number of pre-built graph algorithms.
It is not open-source and its back-end relies on internal queries written in \texttt{C++} to work with a global address space using multiple processes on multiple compute nodes to share data and synchronize operations.
This product being both proprietary and reliant on custom hardware has the consequence of not being so widespread.
However, its results and special-purpose architecture make it a competitive platform which harnessed innovation in design as a graph database.

\item \texttt{On\-to\-text} \texttt{Gra\-ph\-DB}~\cite{ontotext_graphdb} is a graph database focused on \texttt{RDF} data and offering ACID transaction properties.  
It comes in three editions: free which is used for smaller projects and for testing and is only able to execute at most two concurrent queries; standard which can load and query statements at scale; enterprise edition which offers horizontal scalability and other features.
It supports \texttt{SPAR\-QL} and offers a \texttt{Ja\-va} programming API. 
We did not find its source code available.

\end{itemize}

The following support at least both the property graph model and/or \texttt{RDF} explicitly plus other data models (e.g., at least any of: document collections, relational model, object model):

\begin{itemize}


\item \texttt{Amazon Neptune}~\cite{bebee2018amazon} is a managed proprietary service (freeing the user from having to focus on management tasks, provisioning, patching, etc.) that is ACID-compliant and focused on highly-connected datasets and among its use cases are recommendation engines, fraud detection and drug discovery, among others.
Its implementation language and internal graph representation have not been disclosed and it supports both the property graph and \texttt{RDF} models, offering \texttt{Grem\-lin} and \texttt{SPAR\-QL} to query them.
As a full-fledged commercial product, it also has many features related to backup, replicas, security and management tasks, using product features such as Amazon's \texttt{S3}, \texttt{EC2} and \texttt{CloudWatch} to offer scalability.
Usage samples available online~\cite{neptune_github}.

\item \texttt{An\-zo\-Gra\-ph} \texttt{DB}~\cite{anzodb} (previously \texttt{SPARQLVerse}) is a proprietary database built to enable \texttt{RDF} with \texttt{SPAR\-QL} and the property graph with \texttt{Cy\-pher} queries to analyze big graphs (trillions of relationships) and it has \texttt{Ja\-va} and \texttt{C++} APIs to create functions, aggregates and services.
It supports ACID transactions and also supports an \texttt{RFDF+} inference engine following W3C standards and uses compressed in-memory and on disk storage of data.
This database is described as beyond a transaction-oriented database and as a \texttt{Gra\-ph On\-li\-ne A\-na\-ly\-tics Pro\-ce\-s\-s\-ing\-(GO\-LAP)} database, enabling interactive view, analysis and update of graph data, in a way similar to the interactive capabilities of the \texttt{Neo4j} database.
It comes as a single-machine (and memory usage limitations) free edition and an enterprise edition which supports unlimited cluster size, both supporting commercial and non-commercial use.
It supports third-party visualization tools.
We did not find details on its internal data structures nor source code online.

\item \texttt{A\-ran\-go\-DB} is an open-source~\cite{arangodb_github} mul\-ti\--threa\-ded database with support for graph storage (as well as key/value pairs and documents) that is available in both a proprietary license and the \texttt{A\-pa\-che} \texttt{Li\-cen\-se} \texttt{2.0}.
It is written in \texttt{Ja\-va\-Scri\-pt} from the browser to the back-end and all data is stored as \texttt{JSON} documents.
\texttt{A\-ran\-go\-DB} provides a storage engine for mostly in-memory operations and an alternative storage engine based on \texttt{Ro\-cks\-DB}, enabling datasets that are much bigger than RAM.
It guarantees ACID transactions for multi-do\-cu\-ment and m\-ul\-ti\--co\-l\-lec\-tion queries in a single instance and for single-document operations in cluster mode.
Replication and sharding are offered, allowing users to set up the database in a master-slave configuration or to spread bigger datasets across multiple servers.
It exposes a \texttt{Pre\-gel}-like API to express graph algorithms (implying access to the stored data in the database), has a custom \texttt{SQL}-like query language called \texttt{AQL (ArangoDB Query Language)} and includes a built-in graph explorer.

\item \texttt{IBM System G}~\cite{ibm_suite_g} more than a proprietary graph database, is a complete suite of functionalities, able to support the property graph (with \texttt{Grem\-lin}) as well as \texttt{RDF} (though we did not find comments on \texttt{RDF}-specific query languages).
It is comprised of proprietary components as well as open-source and comes with visual query capabilities, providing visual feedback into query building and result analysis to ease the debugging process.
ACID transactions are supported and the graph is represented in its native store with a data structure similar to compressed sparse vectors, using offsets to delimit, for each graph element, the latest and earliest temporal information of the element.

\item \texttt{O\-rient\-DB}~\cite{tesoriero2013getting} is a distributed (pro\-per\-ty mo\-del) graph database that supports \texttt{Tin\-ker\-Pop} and functions both as a graph database and \texttt{No\-SQL} document database as well, with a Community Edition licensed~\cite{orientdb_github} under \texttt{A\-pa\-che} \texttt{Li\-cen\-se} \texttt{2.0} and a commercial edition.
There are drivers supporting \texttt{O\-rient\-DB} at least in the following languages: \texttt{Clojure}, \texttt{Go}, \texttt{Ja\-va}, \texttt{Ja\-va\-Scri\-pt}, \texttt{.NET}, \texttt{Node.js}, \texttt{PHP}, \texttt{Py\-thon}, \texttt{R}, \texttt{Ru\-by} and \texttt{Scala}.
It supports sharding (horizontal scaling), has ACID support and offers an adapted \texttt{SQL} for querying.

\item \texttt{O\-ra\-cle Spa\-tial and Gra\-ph (O\-S\-G)}~\cite{oracledb_graph_features} is a long-standing commercial product which has spatial and graph capabilities, among which the property graph model using \texttt{PGQL} and the \texttt{RDF} model with \texttt{SPAR\-QL}. 
It also supports a feature-rich studio with notebook interpreters, shell user-interface and graph vi\-sua\-li\-za\-tion.
The\-re are different \texttt{Ja\-va} APIs, one for the \texttt{O\-ra\-cle Spa\-tial and Gra\-ph Pro\-per\-ty Gra\-ph}, another for \texttt{Tin\-ker\-Pop Blue\-prin\-ts} and \texttt{Da\-ta\-ba\-se Pro\-per\-ty Gra\-ph}.

\item \texttt{Star\-dog}~\cite{stardog} is a proprietary (with a limited-time free trial version and a paid Enterprise license) knowledge graph database with a graph model based on \texttt{RDF} and extensions to support the property graph model.
It is horizontally-scalable, supports ACID operations, \texttt{Gra\-ph\-QL}, \texttt{Grem\-lin} and \texttt{SPAR\-QL} for querying and introspection and may be programmed in \texttt{Clojure}, \texttt{Groovy}, \texttt{Ja\-va}, \texttt{Ja\-va\-Scri\-pt}, \texttt{.NET} and \texttt{Spring}.
For exploration, it features \texttt{Stardog Studio}.

\item \texttt{Vir\-tu\-oso}~\cite{erling2012virtuoso} is a multi-model database management system that supports relational as well as property graphs and has an open-source~\cite{virtuoso_github} edition under the \texttt{GPLv2} license as well as a proprietary enterprise edition.
At least the following programming languages are supported: \texttt{C\-/C++}, \texttt{C\#}, \texttt{Ja\-va}, \texttt{Ja\-va\-Scri\-pt}, \texttt{.NET}, \texttt{PHP}, \texttt{Py\-thon}, \texttt{Ru\-by} and \texttt{Vi\-sual} \texttt{Ba\-sic}.
It supports horizontal scaling, has functionalities for interactive data exploration and supports \texttt{SPAR\-QL}.

\end{itemize}

We lastly note the following databases that have been used to represent graphs, though not having an explicit description of supporting the property graph model or \texttt{RDF}:

\begin{itemize}

\item \texttt{Azure Cosmos DB}~\cite{paz2018introduction} is a co\-m\-mer\-cial da\-ta\-ba\-se solution that is multi-model, glo\-ba\-l\-ly\--dis\-tri\-bu\-ted, schema-agnostic, ho\-ri\-zon\-tal\-ly\--sca\-la\-ble and fully supports ACID.
It is classified as a \texttt{No\-SQL} da\-ta\-ba\-se, but the multi-model API is a relevant offering, for it can expose stored data for example as table rows (\texttt{Ca\-s\-san\-dra}), collections (\texttt{Mon\-go\-DB}) and most importantly as graphs (\texttt{Grem\-lin}).
It has connectors for \texttt{Ja\-va}, \texttt{.NET}, \texttt{Py\-thon} and \texttt{Xa\-ma\-rin}.
It is also a fully-managed service with scalability, freeing developer resources from topics such as data center deployments, software upgrades and other operations.
An online repository of source code information is available~\cite{cosmos_github}. 

\item \texttt{Fau\-na\-DB}~\cite{faunadb} is a distributed database platform targeting the modern cloud and container-centric environments. 
It has a custom \texttt{Fau\-na Que\-ry Lan\-gua\-ge} (\texttt{FQL}) which operates on schema types such as documents, collections, indices, sets and databases.
This language can be accessed through drivers in languages such as \texttt{Android}, \texttt{C\#}, \texttt{Go}, \texttt{Ja\-va}, \texttt{Ja\-va\-Scri\-pt}, \texttt{Py\-thon}, \texttt{Ru\-by}, \texttt{Scala} and \texttt{Swift}.
\texttt{Fau\-na\-DB} supports concurrency, ACID transactions and offers a RESTful HTTP API.

\item \texttt{Google Cayley} is an open-source~\cite{cayley_github} da\-ta\-ba\-se behind Google's Knowledge Graph, having been tes\-ted at least since 2014 (and it is the spiritual successor to \texttt{graphd}~\cite{Meyer:2010:OST:1807167.1807283}).
It is a community-driven databa\-se written in \texttt{Go}, including a RE\-PL, a RESTful API, a query editor and visualizer.
It supports \texttt{Giz\-mo} (query language inspired by \texttt{Grem\-lin}) and \texttt{Gra\-ph\-QL}.
\texttt{Cay\-ley} supports multiple storage back-end such as \texttt{Le\-vel\-DB}, \texttt{Bolt}, \texttt{Post\-gre\-SQL}, \texttt{Mon\-go\-DB} (distributed stores) and also an ephe\-mer\-al in\--me\-mo\-ry storage.
The ability to support ACID transactions is delegated to the underlying storage back-end.
\texttt{Cay\-ley} being distributed depends on the underlying storage being dis\-tri\-bu\-ted as well.
Also in active development as of 2019.

\item \texttt{Hy\-per\-Gra\-ph\-DB}~\cite{iordanov2010hypergraphdb} is as an open-source general purpose data storage mechanism.
It is used to store hypergraphs (a graph generalization where an edge can join any number of vertices). 
The low-level storage is based on \texttt{Ber\-ke\-ley\-DB} and is implemented in \texttt{Ja\-va}. 
Despite the description on the website mentioning distribution, the support for either distributed sharding or distributed replication is not supported in its current implementation~\cite{hypergraphdb_not_distributed} and we did not find mentions of ACID transaction support.

\item \texttt{O\-bjecti\-vi\-ty/DB}~\cite{objectivitydb} is the database technology powering the massively scalable graph software platform \texttt{ThingSpan} (formerly known as \texttt{InfiniteGraph}).  
It is a fully-distributed database (able to scale horizontally) offering APIs in \texttt{C++}, \texttt{C\#}, \texttt{Ja\-va} and \texttt{Py\-thon}.
\texttt{O\-bjecti\-vi\-ty/DB} is described as a distributed object database, supporting many data models (a\-mong which highly complex and inter-related data).
We  did not find any information about it supporting graph-specific query languages, and licensing is defined on a case-by-case basis.

\end{itemize}

\begin{table*}[h!] 
	\scriptsize
	\centering
	\setlength\tabcolsep{1.5pt} 
	\begin{tabularx}{12.5cm}{p{2.7cm} |p{0.166cm}| p{0.166cm}| p{0.166cm}| c c c c c } 
		\toprule 
		
		\emph{Database} & \rotatebox[origin=c]{90}{\emph{ACID}}& \rotatebox[origin=c]{90}{\emph{Visual}} & \rotatebox[origin=c]{90}{\emph{Scale}}   & \emph{Models} & \emph{Languages} & \emph{GQLs} &  \emph{License} & \emph{Limits} \\\midrule
		
		\texttt{Ali\-ba\-ba} \texttt{GDB}~\cite{alibaba_graphdb}    & \blackcirclemarker  &   & \blackcirclemarker  & \texttt{PG} & \texttt{Go}, \texttt{Ja\-va}, \texttt{.NET}, \texttt{Node.js},  \texttt{Py\-thon} & \texttt{Grem\-lin} & \textit{Paid} & \textit{N/A} \\ 
		\texttt{Chro\-no\-Gra\-ph}~\cite{haeusler2017chronograph,chronograph_github}    & \blackcirclemarker  &  &  & \texttt{PG} & \texttt{Ja\-va} & \texttt{Grem\-lin} & \texttt{aGPL v3} & \textit{N/A}  \\ 
		\texttt{DSE}~\cite{datastax} & \blackcirclemarker   & \blackcirclemarker   & \blackcirclemarker   & \texttt{PG} & \texttt{Ja\-va} & \texttt{Grem\-lin} & \textit{Paid} & \textit{N/A} \\ 
		\texttt{D\-gra\-ph}~\cite{dgraph_graphql,dgraph_github} & \blackcirclemarker  &  & \blackcirclemarker  & \texttt{PG} & \texttt{Ja\-va}, \texttt{Ja\-va\-Scri\-pt}, \texttt{Py\-thon} & \textit{GraphQL} & \texttt{AL 2.0} & \textit{Admin} \\
		\texttt{Graph\-flow}\cite{kankanamge2017graphflow,10.14778/3342263.3342643,mhedhbi2020a+} &   &   &   & \texttt{PG} & \texttt{Ja\-va} & \textit{Cypher} &  \texttt{AL 2.0}  & \textit{N/A} \\ 
		\texttt{Ja\-nus\-Gra\-ph}~\cite{janus,janus_github} & \blackcirclemarker  &   & \blackcirclemarker & \texttt{PG} & \makecell{\texttt{Elixir}, \texttt{Go}, \texttt{Ja\-va}, \texttt{.NET}, \texttt{PHP},\\ \texttt{Py\-thon}, \texttt{Ru\-by}, \texttt{Scala}} & \texttt{Grem\-lin} & \texttt{AL 2.0} & \textit{N/A} \\ 
		\texttt{Ne\-bu\-la} \texttt{Gra\-ph}~\cite{nebulagraph_github} & \blackcirclemarker  & \blackcirclemarker  & \blackcirclemarker  & \texttt{PG} & \texttt{Go}, \texttt{Ja\-va}, \texttt{Py\-thon} & \textit{nGQL} & \texttt{AL 2.0, CC 1.0} & \textit{N/A} \\
		\texttt{Neo4j}~\cite{Webber:2012:PIN:2384716.2384777,neo4j_github} & \blackcirclemarker  & \blackcirclemarker  &  & \texttt{PG} & \makecell{ \texttt{C\-/C++}, \texttt{Clojure}, \texttt{Go}, \texttt{Haskell}, \\\texttt{Ja\-va}, \texttt{Ja\-va\-Scri\-pt}, \texttt{.NET}, \texttt{Perl},\\ \texttt{PHP}, \texttt{Py\-thon}, \texttt{R}, \texttt{Ru\-by}} & \texttt{Cy\-pher} & \texttt{GPL v3} & \textit{Size} \\ 
		\texttt{Re\-dis\-Gra\-ph}~\cite{cailliau2019redisgraph,redis_github} & &  &  & \texttt{PG} & \makecell{\texttt{Elixir}, \texttt{Go}, \texttt{Ja\-va}, \texttt{Ja\-va\-Scri\-pt},\\ \texttt{PHP}, \texttt{Py\-thon}, \texttt{Ru\-by}, \texttt{Rust}} & \textit{Cypher} & \textit{Custom} & \textit{N/A}\\ 
		\texttt{Hana} \cite{rudolf2013graph,hwang2018graph,hana_github} & \blackcirclemarker  & \blackcirclemarker  & \blackcirclemarker  & \texttt{PG} & \textit{N/A} & \texttt{Cy\-pher} & \textit{Paid} & \textit{N/A} \\ 
		\texttt{Spark\-see}~\cite{martinez2011dex,sparksee_former_dex} & \blackcirclemarker  &   &  & \texttt{PG} & \makecell{\texttt{C++}, \texttt{.NET}, \texttt{Ja\-va}, \\\texttt{Objective-C}, \texttt{Py\-thon}} & \textit{N/A} & \textit{Custom} & \textit{Size} \\ 
		\texttt{Ti\-ger\-Gra\-ph} \cite{deutsch2019tigergraph,tigergraph} & \blackcirclemarker  &  & \blackcirclemarker & \texttt{PG} & \textit{N/A} & \textit{GSQL} & \textit{Custom} & \textit{Scale} \\ 
		\texttt{Wea\-ver}~\cite{DBLP:journals/pvldb/DubeyHES16,weaver_github} & \blackcirclemarker  &  & \blackcirclemarker  & \texttt{PG} & \texttt{C++}, \texttt{Py\-thon} & \textit{N/A}  &  \textit{Custom} & \textit{N/A} \\
		
		\midrule 
		
		\texttt{Alle\-gro\-Gra\-ph}~\cite{allegrograph}    & \blackcirclemarker  &  & \blackcirclemarker & \texttt{RDF} & \makecell{\texttt{C\#}, \texttt{C}, \texttt{Common Lisp}, \texttt{Clojure}, \\\texttt{Ja\-va}, \texttt{Perl}, \texttt{Py\-thon}} &  \texttt{SPAR\-QL} & \texttt{EPL v1}  &  \textit{Size}   \\ 
		\texttt{Bla\-ze\-Gra\-ph}~\cite{blazegraph}    & \blackcirclemarker  &  &  & \makecell{\texttt{PG}, \texttt{RDF}, \textit{O}} & \makecell{\texttt{.NET}, \texttt{Py\-thon}} & \makecell{ \texttt{Grem\-lin},\\ \texttt{SPAR\-QL}} &  \texttt{GPL v2} &   \textit{N/A} \\ 
		\texttt{Bri\-ght\-star\-DB}~\cite{brightstardb_github}    & \blackcirclemarker  &  &  & \texttt{RDF} & \texttt{.NET} & \texttt{SPAR\-QL} & \texttt{MIT} & \textit{N/A}   \\ 
		\texttt{Cray} \texttt{Gra\-ph} \texttt{En\-gi\-ne}~\cite{rickett2018loading} & \blackcirclemarker  &  & \blackcirclemarker  & \texttt{RDF} & \makecell{\texttt{Ja\-va}, \texttt{Py\-thon}}   & \texttt{SPAR\-QL} & \textit{Paid} & \textit{N/A}  \\
		\texttt{On\-to\-text} \texttt{Gra\-ph\-DB}~\cite{ontotext_graphdb} & \blackcirclemarker  & \blackcirclemarker & \blackcirclemarker  & \texttt{RDF}  & \texttt{Ja\-va}, \texttt{Ja\-va\-Scri\-pt} & \texttt{SPAR\-QL} & \textit{Custom} & \textit{Scale}  \\
		
		\midrule 

		\texttt{Nep\-tu\-ne}~\cite{bebee2018amazon,neptune_github}    & \blackcirclemarker &   & \blackcirclemarker  & \texttt{PG},\texttt{RDF} & \textit{Managed} & \makecell{\texttt{Grem\-lin},\\ \texttt{SPAR\-QL}} & \textit{Paid} & \textit{N/A} \\ 
		\texttt{An\-zo\-Gra\-ph} \texttt{DB}~\cite{anzodb}    & \blackcirclemarker  & \blackcirclemarker & \blackcirclemarker & \texttt{PG},\texttt{RDF} & \texttt{Ja\-va},\texttt{C++} &  \makecell{\texttt{Grem\-lin},\\ \texttt{SPAR\-QL}} & \textit{Custom} & \textit{Scale}  \\ 
		\texttt{A\-ran\-go\-DB}~\cite{arangodb_github}    & \blackcirclemarker  & \blackcirclemarker & \blackcirclemarker  & \texttt{PG} & \makecell{\texttt{Go}, \texttt{Ja\-va}, \texttt{Ja\-va\-Scri\-pt}, \texttt{PHP}} & \textit{AQL} &  \texttt{AL 2.0} & \textit{Admin}  \\ 
		\texttt{IBM System G}~\cite{ibm_suite_g}    & \blackcirclemarker  & \blackcirclemarker  & \blackcirclemarker  & \texttt{PG}, \texttt{RDF} & \textit{Managed} & \texttt{Grem\-lin} & \textit{Paid} & \textit{N/A}  \\ 
		\texttt{O\-rient\-DB}~\cite{tesoriero2013getting,orientdb_github}    & \blackcirclemarker  & \blackcirclemarker & \blackcirclemarker  & \texttt{PG}, \textit{O} & \makecell{\texttt{Clojure}, \texttt{Go}, \texttt{Ja\-va}, \\\texttt{Ja\-va\-Scri\-pt},\texttt{.NET}, \texttt{Node.js}, \texttt{PHP},\\ \texttt{Py\-thon}, \texttt{R}, \texttt{Ru\-by}, \texttt{Scala}} & 6 & \texttt{AL 2.0} & \textit{Visual}  \\ 
		\texttt{OSG}~\cite{oracledb_graph_features}    & \blackcirclemarker  & \blackcirclemarker & \blackcirclemarker  & \texttt{PG}, \texttt{RDF} & \texttt{Ja\-va} & \makecell{\texttt{PGQL},\\ \texttt{SPAR\-QL}} & \textit{Paid} & \textit{N/A} \\
		\texttt{Star\-dog}~\cite{stardog}    & \blackcirclemarker  & \blackcirclemarker  & \blackcirclemarker  & \texttt{RDF} & \makecell{\texttt{Clojure}, \texttt{Groovy}, \texttt{Ja\-va},\\ \texttt{Ja\-va\-Scri\-pt}, \texttt{.NET}} & \makecell{\texttt{Gra\-ph\-QL}, \\\texttt{Grem\-lin}, \\\texttt{SPAR\-QL}} & \textit{Paid} & \textit{Trial}  \\
		\texttt{Vir\-tu\-oso}~\cite{erling2012virtuoso,virtuoso_github}    & \blackcirclemarker  & & \blackcirclemarker  & \texttt{PG}, \textit{O} & \makecell{\texttt{C\-/C++}, \texttt{C\#}, \texttt{Ja\-va}, \texttt{Ja\-va\-Scri\-pt},\\ \texttt{.NET}, \texttt{PHP}, \texttt{Py\-thon},\\ \texttt{Ru\-by}, \texttt{Vi\-sual} \texttt{Ba\-sic}} & \texttt{SPAR\-QL} & \texttt{GPL v2} & \textit{Admin} \\
		
		\midrule 
		
		\texttt{Cos\-mos} \texttt{DB}~\cite{paz2018introduction,cosmos_github}    & \blackcirclemarker &  & \blackcirclemarker  & \textit{MM} & \texttt{Grem\-lin} & \textit{Managed} & \textit{Paid} & \textit{N/A} \\
		\texttt{Fau\-na\-DB}~\cite{faunadb}    & \blackcirclemarker  &  & \blackcirclemarker & \textit{MM} & \makecell{\texttt{Android}, \texttt{C\#}, \texttt{Go}, \texttt{Ja\-va},\\ \texttt{Ja\-va\-Scri\-pt}, \texttt{Py\-thon}, \texttt{Ru\-by},\\ \texttt{Scala}, \texttt{Swift}} & \textit{FQL} & \textit{Custom} & \textit{Quota}  \\
		\texttt{Cay\-ley}~\cite{cayley_github,Meyer:2010:OST:1807167.1807283}    & \blackcirclemarker  & \blackcirclemarker  & \blackcirclemarker  & \texttt{RDF} & \textit{N/A} & \makecell{\texttt{Gizmo},\\\texttt{Gra\-ph\-QL}} & \texttt{AL 2.0} & \textit{N/A}  \\
		\texttt{Hy\-per\-Gra\-ph\-DB}~\cite{iordanov2010hypergraphdb}    & & & & \textit{HG} & \texttt{Ja\-va} & \textit{N/A} & \texttt{AL 2.0} & \textit{N/A} \\
		\texttt{O\-bjecti\-vi\-ty/DB}~\cite{objectivitydb}    & \blackcirclemarker  & & \blackcirclemarker  & \textit{O}  & \makecell{\texttt{C++}, \texttt{C\#}, \texttt{Ja\-va}, \texttt{Py\-thon}} & \textit{N/A} & \textit{Paid} & \textit{N/A} \\

		\bottomrule 
	\end{tabularx}
	\caption{Summary of graph database distinctive features. Circle \blackcirclemarker\ on the \emph{ACID}, \emph{Visual} and \emph{Scale} (updating a graph that is distributed) columns indicate they are present. \emph{Models} lists the supported data models: property graph model (\texttt{PG}), \texttt{RDF}, multi-model (\textit{MM} or others (\textit{O}). \emph{Languages} lists the programming languages that interface with the database, either directly or by being supported by the supported interfaces (e.g., \texttt{Tin\-ker\-Pop} drivers). \emph{GQLs} lists only directly supported graph query languages (and shown in italic if they are not standard) - those supported using compatibility tools as seen previously are not included. \emph{License} lists the licenses of the open-source project or of the free edition of a commercial product: \texttt{AL 2.0} is \texttt{A\-pa\-che} \texttt{Li\-cen\-se} \texttt{2.0}, \texttt{CC 1.0} is \texttt{Commons Clause 1.0}, \texttt{aGPL v3} is the \texttt{Affero General Public License 3}, \texttt{(GPL) v3} is \texttt{GNU General Public License (GPL) v3}. \emph{Limits} covers free version limitations in commercial products (\textit{N/A} if it does not apply). If a database has \textit{Custom} license and limits are not \textit{N/A}, it means it has a non-standard free license.}
	\label{table:graph_database_features}
\end{table*}

Other notable mentions include the \texttt{OQGRAPH}~\cite{mariadb_oqgraph} graph storage engine of \texttt{MariaDB}, developed to handle hierarchies and vertices with many connections and intended for retrieving hierarchical information such as graphs, routes and social relationships in \texttt{SQL}. 
We note the following resources for further deepening of the aspects involved in the evolution of the study of graph databases~\cite{jin2010gblender,buerli2012current,shimpi2012overview,Robinson:2013:GD:2556013,kolomivcenko2013experimental,graphDBs,de2014model,henderson2014system,robinson2015graph}, covering aspects such as the internal graph representations, experimental comparisons, principles for querying and extracting information from the graph and designing graph da\-ta\-ba\-ses to make use of distributed infrastructures.

\section{Conclusion}\label{sec:remarks}

This survey explores different aspects of the graph processing landscape and highlights vectors of research.
We cover dimensions that enable the classification of graph processing systems according to the mutability of data (dynamism~\cite{besta2019practice} and its modalities), the nature of the tasks (workloads where the focus may be efficient storage~\cite{oltp_oracle} or swift computation~\cite{4781105} over transient data) and how the data is associated to different computing agents (e.g., distributed via partitioning~\cite{soudani2019investigation} to threads in a CPU, CPUs in a machine, machines in a cluster).
Each of these dimensions constitutes a different branch of the study of graph processing, and herein we group their recent literature surveys and draw on their relationships.
On drawing a line between graph processing systems and those that also focus on the storage, the graph databases, we found most commercial graph solutions to fall on the category of graph database. 
Graph databases, along the last decade, have continued to refine their efficiency in executing traversals and global graph algorithms over the graph representation stored in the database. 
We consider that a novel approach to extracting value from graph-based data will include the use of graph-aware data compression techniques on scalable distributed systems, potentially breaking the abstraction that these systems establish between the high-level graph data representations and the lower-level data distribution and transmission. 
We observe that the architecture of systems targeting graphs depend on how generic is the graph processing desired to be. 
Generic dataflow processing systems offer abstractions over their basic computational primitives in order to represent and process graphs, but in exchange abdicate from fine-tuning and graph-aware optimizations.

As part of our exhaustive analysis of existing contributions of different domains in the state-of-the-art of graph processing and storage, we provide direct links to source code repositories such as GitHub whenever they were available. 
Should the reader wish to delve into the implementation of a given contribution, a link to the contribution's source code repository is to be found as part of the bibliography.
We provide these so that other researchers and developers may look into them without need to engage in error-prone searches looking for up-to-date documentation and source-code.

This systematic analysis fosters some additional comments regarding data processing.
Data is abundant, big and evolving, and paradigms such as edge computing and the evolution of the In\-ter\-net\--of\--Thin\-gs come together to reshape our relationship with data.
With an increase in \textit{smart} devices and computational capabilities becoming more ubiquitous for example in daily objects such as vehicles and smart homes, new graphs of data mapping interaction and purpose become available.
This implies a continuous trend in the increasing size of data.
At the same time, the dimension of dynamism (spread across the types we enumerate in this document) gains renewed importance as we move to a faster and ever-connected world.
With the advent of 5G technologies and the alternative possibilities of \textit{space internet} (among the private initiatives we count SpaceX's Starlink, Jeff Bezos' Blue Origin and the late Steve Jobs' vision for an always-connected smartphone) becoming a closer reality, the temporal aspect will become even more granular. 

One would not be wrong to speculate that we will have more devices which will generate data more frequently.
In such a world, the graph processing dimensions we enumerate in this document will play a relevant role in building systems to handle these changing scenarios.

\textbf{Funding.} This work was partly supported by national funds through FCT -- Fundação para a Ciência e Tecnologia, under projects PTDC/CPO\--CPO/28495/2017, P\-T\-DC\-/CCI-BIO/29676/2017, P\-TDC/EEI-COM/30644/2017 and UIDB/50021/2020.


\scriptsize{
\bibliographystyle{abbrv}
\bibliography{full}
}


\end{document}